%% file: root.tex
\documentclass[letterpaper, 10 pt, conference]{ieeeconf}  

\IEEEoverridecommandlockouts                              

\usepackage{graphicx}
\usepackage{textcomp}
\usepackage{amsmath, amssymb}
\usepackage{nicefrac}       
\usepackage{comment}
\usepackage{siunitx}
\usepackage{aligned-overset}
\usepackage{algorithm}
\usepackage[noend]{algpseudocode}
\usepackage{xcolor}
\usepackage{tikz}
\usepackage{cite}
\usepackage{pgfplots}
\usetikzlibrary{positioning}
\let\labelindent\relax
\usepackage[shortlabels]{enumitem}
\usepackage{pgfplots}
\usepackage{wrapfig}
\usetikzlibrary{graphs}
\usepgfplotslibrary{groupplots}
\usepackage{float}
\usepackage{comment}

\input{macros}
\usepackage[most]{tcolorbox}
\usepackage{xcolor}
\usepackage{lipsum} 
\usepackage[framemethod=tikz]{mdframed}
\usepackage[colorlinks=true, linkcolor=blue, citecolor=blue, urlcolor=blue]{hyperref}

\title{\LARGE \bf
Towards safe control parameter tuning in\\ distributed multi-agent systems}

\author{Abdullah Tokmak$^{1}$, Thomas B.\ Schön$^{2}$, and Dominik Baumann$^{1}$
\thanks{*This research was partially supported by \emph{Kjell och Märta Beijer Foundation}.
}
\thanks{$^{1}$
Cyber-physical Systems Group, Aalto University, Espoo, Finland {\tt\small firstname.lastname@aalto.fi}}%
\thanks{$^{2}$
Department of Information Technology,
        Uppsala University, Uppsala, Sweden
        {\tt\small thomas.schon@it.uu.se}}%
}

\usepackage{fancyhdr}
\fancypagestyle{firstpage}{%
  \fancyhf{} 
  \fancyfoot[L]{\normalfont \sffamily \scriptsize \lowernote}
}
\fancypagestyle{plain}{%
  \fancyhf{}
  \fancyfoot[L]{\normalfont \sffamily \scriptsize \lowernote}
}

\newcommand{\lowernote}{
\textbf{Accepted final version. \textcopyright\; 2025
IEEE.}
To be presented at the Conference on Decision and Control (CDC), 2025.
Personal use of this material is permitted. Permission from IEEE must be obtained for all other uses, in any current or future media,
including reprinting/republishing this material for advertising or promotional purposes, creating new collective works, for resale or redistribution to servers
or lists, or reuse of any copyrighted component of this work in other works.
}
\begin{document}
\newmdenv[innerlinewidth=0.5pt, roundcorner=4pt,linecolor=aaltoBlue,innerleftmargin=6pt,
backgroundcolor=aaltoBlue!10,
innerrightmargin=6pt,innertopmargin=6pt,innerbottommargin=6pt]{mybox}

\maketitle
\thispagestyle{firstpage}


\begin{abstract}

Many safety-critical real-world problems, such as autonomous driving and collaborative robots, are of a distributed multi-agent nature. 
To optimize the performance of these systems while ensuring safety, we can cast them as distributed optimization problems, where each agent aims to optimize their parameters to maximize a coupled reward function subject to coupled constraints.
Prior work either studies a centralized setting, does not consider safety, or struggles with sample efficiency.
Since we require sample efficiency and work with unknown and nonconvex rewards and constraints, we solve this optimization problem using safe Bayesian optimization with Gaussian process regression.
Moreover, we consider nearest-neighbor communication between the agents. 
To capture the behavior of non-neighboring agents, we reformulate the static global optimization problem as a time-varying local optimization problem for each agent, essentially introducing time as a latent variable.
To this end, we propose a custom spatio-temporal kernel to integrate prior knowledge.
We show the successful deployment of our algorithm in simulations.
\end{abstract}

\input{Sections/introduction.tex}

\input{Sections/problem_setting}

\input{Sections/algorithm.tex}

\input{Sections/experiments.tex}

\input{Sections/conclusions.tex}








\bibliographystyle{IEEEtran}
\bibliography{IEEEabrv,mybibfile}

\end{document}

%% file: macros.tex
\newcommand{\safeopt}{\textsc{SafeOpt}}

\newcommand{\ie}{i\/.\/e\/.,\/~}
\newcommand{\eg}{e\/.\/g\/.,\/~}

\newtheorem{proposition}{\bf Proposition}

\newcommand{\domain}{\mathcal A}

\DeclareMathOperator*{\argmax}{arg\,max\,}

  \definecolor{aaltoRed}{RGB}{239,51,64}%
  \definecolor{aaltoBlue}{RGB}{0,94,184}%

\DeclareMathOperator*{\R}{\mathbb{R}}

\newcommand{\A}[1]{{({#1})}}
\renewcommand{\a}{\mathbf{a}}
\newcommand{\N}[1]{{(\mathcal{N}^{#1}_+)}}

\newcounter{tool}
\renewcommand{\thetool}{(T\arabic{tool})}
\newcommand{\tool}[2]{%
  \refstepcounter{tool}%
  \label{#1}%
  \textbf{\thetool} \emph{#2}%
}

\definecolor{magenta}{RGB}{255,0,255}
\definecolor{lightgreen}{RGB}{76, 200, 76}
\definecolor{deeporange}{RGB}{204, 112, 0}
\newcommand{\lline}{$\ell.$}

%% file: Sections/introduction.tex
\section{INTRODUCTION}

Collaborative robots, autonomous driving, swarm robotics in warehouses, and many other real-world problems can be realized as distributed multi-agent systems (MAS). 
Distributed MAS do not have a centralized node that collects the information about the environment and coordinates other agents.
Therefore, agents in distributed settings must optimize their control policies \emph{individually} to reach a collaborative goal.
When environments and system dynamics are unknown, learning-based approaches offer a constructive tool to learn control policies from data~\cite{marl-book}. 
When parameterizing the control policies, learning a desirable behavior can be seen as tuning policy parameters to maximize a reward function that quantifies their performance.
While tuning the control parameters, we require~\emph{(i)} sample efficiency, since every sample corresponds to a real-world experiment; and~\emph{(ii)} safety guarantees~\cite{brunke2022safe, garg2024learning}, since applications like autonomous driving are safety-critical and policy failures can endanger nearby personnel or damage hardware.
Although reinforcement learning (RL)~\cite{sutton2018reinforcement} is widely used for policy learning, it struggles with sample efficiency and safety guarantees.
In contrast, 
Bayesian optimization (BO)~\cite{garnett2023bayesian, frazier2018tutorial} with Gaussian process (GP) regression~\cite{Rasmussen2006Gaussian} is an effective method for data-efficient policy learning in real-world systems~\cite{marco2016automatic}.
Moreover, safe BO algorithms additionally provide probabilistic safety guarantees~\cite{sui2015safe}.
However, applying safe BO algorithms to distributed MAS is an open challenge, specifically due to the coupled nature of the rewards and constraints.
In particular, the parameter choice of any agent influences the reward value that all others receive.
However, in general, agents are not aware of the policy parameter choices of each other.
Hence, every agent has unobserved sub-spaces.

%

\subsubsection*{Contribution}
In this work, we present a BO algorithm that safely tunes control parameters of distributed MAS with nearest-neighbor communication.
Specifically, we make the following contributions:
\begin{itemize}
    \item We consider the behavior of non-neighboring agents by introducing time as a latent variable, thereby establishing a time-varying interpretation of the static global reward function.
    \item We develop a custom spatio-temporal kernel to model the unknown reward function using GPs.
    \item We propose a BO algorithm for safe control parameter tuning in distributed MAS and demonstrate its effectiveness in numerical examples and a vehicle platooning simulation.
\end{itemize}

\section{BACKGROUND AND RELATED WORK}\label{sec:background}

RL is a popular approach when it comes to policy learning in unknown environments, \ie systems with unknown dynamics and unknown reward functions.
There is extensive research on multi-agent reinforcement learning~\cite{marl-book}, also with a focus on distributed MAS~\cite{zhang2018fully}.
However, RL flourishes with big data, which is often infeasible in real-world applications where each data point corresponds to an experiment.

Another approach to parameter optimization in distributed MAS is to use distributed optimization algorithms that \emph{provably} solve constrained optimization problems with coupled objective functions or coupled constraints.
Two examples are \emph{(i)} the Jacobi method~\cite{saad2003iterative}, which is a popular way to solve cost-coupled problems, and~\emph{(ii)} the alternating direction method of multipliers (ADMM)~\cite{boyd2011distributed}, which solves constraint-coupled problems.
Although both methods guarantee convergence, they rely on (strong) convexity of cost and constraint functions, which is rarely the case when tuning control parameters.
Furthermore, akin to vanilla RL, these algorithms are, to a large extent, sample-inefficient, \ie not directly suitable when function evaluations correspond to real-world experiments.

BO excels at sample efficiency for optimizing unknown functions as it formulates the sample acquisition as an optimization problem itself~\cite{srinivas2012information, chowdhury2017kernelized}.
There are extensions of these algorithms that also guarantee safety, \ie the satisfaction of constraints for every function evaluation, with high probability.
The most popular safe BO algorithm is \safeopt~\cite{sui2015safe}, and its numerous modifications and extensions~\cite{tokmak2024pacsbo, tokmak2024safe, berkenkamp2023bayesian, sui2018stagewise}.
Nevertheless, all of the mentioned works are restricted to single-agent settings.
Reference~\cite{prajapat2022near} proposes an extension of a safe BO algorithm~\cite{turchetta2019safe} to the MAS setting, but they do not consider a fully distributed setting. 
Similarly,~\cite{Steinberg2025MFBO} assumes the presence of a central agent and propose a safe BO algorithm in the mean field.

Fully distributed BO algorithms primarily exist to distribute computation across different cores.
Different from parallelized or batch BO algorithms~\cite{kandasamy2018parallelised, ma2023gaussian}, distributed BO algorithms do not require a central node~\cite[Section~4.1]{wang2023recent}.
A distributed BO algorithm based on distributed Thompson sampling is presented in~\cite{zerefa2024distributed}. 
Unlike our contribution,~\cite{zerefa2024distributed} does not consider constraints and assumes that the unknown function can be evaluated independently by each agent.
Hence,  evaluating---and especially modeling---the unknown function does not involve dealing with unobservable sub-spaces.

Unobservable sub-spaces in BO also implicitly arise when dealing with very high-dimensional problems.
To this end,~\cite{song2022monte, zhang2019high, chen2012joint,Sui2025HDSafeBo} learn a latent representation into a lower dimension and apply BO  on the latent space.
One critical challenge with learning latent variable representations is that they require offline data, which is not available in our setting.

%% file: Sections/problem_setting.tex
\section{PROBLEM SETTING AND PRELIMINARIES}\label{sec:problem}
We cast the problem of safely tuning control parameters as an optimization problem, where the control parameters are the decision variables and the reward function is the objective function.

\subsubsection*{Optimization problem} 
We aim to maximize the unknown reward function $f:\domain^N \rightarrow \mathbb{R}$ while guaranteeing safety.
We consider a MAS with~$N$ agents, where each agent~$i$ has $n$-dimensional control parameters $a_t^{(i)}\in\domain_i\subseteq\R^n$, where~$t\geq 1$ is the iteration counter.
To simplify notation, we assume that~$\domain_i\equiv\domain_j$ for all~$i,j\in\{1,\ldots,N\}$ and hence write~$\domain_i=\domain$.
The reward function is coupled, \ie it depends on the control parameters~$\a_t\coloneqq [a_t^{\A{1}}, \ldots, a_t^{\A{N}}]^\top \in \mathbb R^{N\times n}$ of all~$N$ agents.
We introduce the safety threshold~$h\in\mathbb R$ and define safety as only evaluating~$f$ with parameters~$\a_t$ that result in a reward value larger than~$h$.
Hence, the coupled optimization problem is 
\begin{align}\label{eq:global_opt}
    \max_{\a \in \domain^N} f(\a) \quad \text{subject to } f(\a_t) \geq h, \forall t\geq 1.
\end{align}

We solve~\eqref{eq:global_opt} \emph{episodically} using safe BO, performing the optimization once at the end of each episode, \ie after sampling~$f$.
For the function evaluation of~$f(\a_t)$, we conduct an experiment, where each agent~$i$ applies its local control parameter~$a_t^\A{i}$. 
In return, each agent receives the corresponding global reward value~$y_t\coloneqq f(\a_t)+\epsilon_t$, where~$\epsilon_t$ is~$\sigma$-sub-Gaussian measurement noise.
For each agent~$i\in\{1,\ldots,N\}$, we collect the applied parameters until iteration~$t$ in~$a_{1:t}^\A{i}\coloneqq [a_1^\A{i},\ldots,a_t^\A{i}]^\top$ and the corresponding reward values in~$y_{1:t}\coloneqq [y_1,\ldots, y_t]^\top$.

\subsubsection*{MAS} 
\begin{figure}
\begin{tikzpicture}
    \node[circle, draw, fill=aaltoBlue!20] (A) at (0,0) {1};
    \node[circle, draw, fill=aaltoBlue!20] (B) at (2,1) {2};
    \node[circle, draw, fill=aaltoBlue!20] (C) at (4,0) {3};
    \node[circle, draw, fill=aaltoBlue!20] (D) at (6,0) {4};
    \node[circle, draw, fill=aaltoBlue!20] (E) at (8,1) {5};
    \node[circle, draw, fill=aaltoBlue!20] (F) at (8,0) {6};
    
    \draw[-] (A) -- (B);
    \draw[-] (B) -- (C);
    \draw[-] (C) -- (D);
    \draw[-] (D) -- (E);
    \draw[-] (E) -- (F);
\end{tikzpicture}
\caption{Example of the nearest neighbor communication structure.}
\label{fig:communication}
\end{figure}
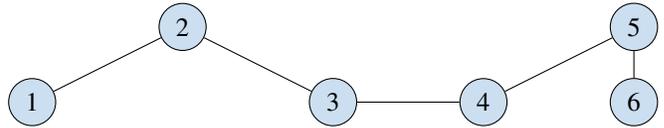

Solving the optimization problem~\eqref{eq:global_opt} is infeasible without communicating the applied control parameters to other agents.
We assume nearest-neighbor communication, represented by the undirected graph in series connection in Figure~\ref{fig:communication}.
If two agents~$i$ and~$j$ can communicate, they have an edge that connects nodes~$i$ and~$j$.
We denote the graph by $G=(V,E)$ with cardinality of the vertices~$\lvert V \rvert=N$ given by the number of agents.
The set of edges is of the form $E \subseteq \{\{i,j\}: i,j \in V\}$.
We define the neighbors of node~$i$ as the nodes that have an edge to node~$i$, \ie $\mathcal N^i \coloneqq \{j\in V: \{i,j\}\in E\}$ and $\mathcal N^{i}_{+}\coloneqq \mathcal N^i \cup \{i\}.$
In this work, we assume that~$\mathcal N^i$ remains constant for all iterations.
We collect the parameters that agent~$i$ and its neighbors~$\mathcal N^i$ applied at iteration~$t$ in a vector~$a_t^\N{i}.$\footnote{%
The dimension of~$a_t^\N{i}$ varies for each agent~$i$ and is given by~$a_t^\N{i}\in\mathbb R^{\lvert \mathcal N_+^i \rvert \times n}$.
For instance, for the third agent illustrated in Figure~\ref{fig:communication}, the parameter vector is given by~$a_t^{\mathcal N_+^3}=[a_t^\A{2}, a_t^\A{3}, a_t^\A{4}]^\top$.
}
Moreover, we restrict communication to first-order communication to enforce a privacy constraint.
That is, each agent~$i$ may only communicate its own parameters~$a_t^\A{i}\in\domain$ to its neighbors~$\mathcal N^i$ without forwarding control information that it receives through the network.

\subsubsection*{Unobservable sub-spaces}
As each agent~$i$ is only aware of~$a_{1:t}^\N{i}$, this introduces an unobservable sub-space spanned by the parameters~$a_{1:t}^\A{j}$ of each non-adjacent agent $j\in V \setminus \mathcal{N}^i_+$.
Therefore, each agent only sees a projection, \ie a surjective and thus non-invertible map~$\pi(\a_t)=a_t^\N{i}$ at each iteration~$t$.
Thus, from the perspective of agent~$i$, the same input~$a_t^\N{i}$ may \emph{deterministically} map to different~$y_t$ values, introducing harsh discontinuities.

%% file: Sections/algorithm.tex
\section{SAFE BO FOR DISTRIBUTED MAS}\label{sec:algorithm}
In this section, we present our algorithm.
Specifically, Section~\ref{sec:time-latent} addresses the problem of unobservable sub-spaces by introducing time as a latent variable, while we define time-varying reward functions for each agent in Section~\ref{sec:time-varying}.
Then, in Section~\ref{sec:spatio-temporal}, we propose a custom spatio-temporal kernel to model the time-varying reward function before explaining our safe BO algorithm in Section~\ref{sec:safe_BO}.

\subsection{Time as a latent variable}\label{sec:time-latent}
One way of dealing with unobservable sub-spaces is to learn a latent representation.
However, since no offline data is available, we cannot follow a classic latent representation approach to encode the parameters of non-neighboring agents.
Instead, we introduce a concrete latent variable with a physical interpretation.
\begin{mybox}
\tool{tool:time-latent}{
Inspired by latent variable approaches, we introduce time as a latent variable.
}
\end{mybox}
First, Tool~\ref{tool:time-latent} enables implicit extrapolation of the behavior of non-adjacent, \ie non-neighboring agents.
Second,~\ref{tool:time-latent} introduces a \emph{well-defined mapping} from the extended input space, \ie the modeled parameters~$a_t^\N{i}$ and the time variable~$t$, to the output space~$y_{1:t}$ for every agent~$i\in V$ and for all iterations~$t\geq 1$.
Essentially, even if, \eg $a_t^\N{i} = a_{t+1}^\N{i}$ and $y_{t} \neq y_{t+1}$, the fact that~$t\neq t+1$ naturally ensures that deterministically every input only maps to one function value, which makes learning the function more accessible for each agent~$i\in V$.

\subsection{Time-varying local optimization problems}\label{sec:time-varying}

With time as a latent variable, every agent aims to maximize a time-varying optimization problem that approximates the static optimization problem. 
This allows us to optimize the control parameters in a distributed way without explicitly modeling the parameters of non-neighboring agents.
In particular, each agent~$i \in V$ optimizes
\begin{align}\label{eq:local_opt}
    \max_{a^\N{i}\in\mathcal A^{\lvert \mathcal{N}_+^i \rvert}}  &f_t^\A{i}(a^\N{i}, t) \\ \text{subject to } &f_t^\A{i}(a^\N{i}, t) \geq h \nonumber
\end{align}
in each iteration~$t\geq 1$.

We use the optimization problem~\eqref{eq:local_opt} as an approximation of the intractable optimization problem~\eqref{eq:global_opt}.
Note that~$f_t^\A{i}(a^\N{i}, t)\equiv f(\a_t)$ for all~$i\in V$ and for all~$t\geq 1$.
That is, by introducing the time-varying reward functions~$f_t^\A{i}$, we do \emph{not} introduce a new sampling oracle but continue to receive the corresponding rewards by evaluating the static reward function~$f$.
Instead, from each agent~$i$'s point of view, the reward function is now of the form~$f_t^\A{i}(a^\N{i}, t)$ in lieu of the static global reward function~$f(\a_t)$ that inherently contains unobservable sub-spaces.

We use safe BO to solve~\eqref{eq:local_opt} to exploit its sample efficiency and due to its ability to handle probabilistic constraint satisfaction.
Akin to other safe BO algorithms~\cite{sui2015safe, berkenkamp2023bayesian}, we utilize GPs to model a surrogate of the unknown reward function~$f_t^\A{i}$.
A GP is a stochastic process that is fully characterized by its prior mean and kernel function~$k$.
For each agent~$i\in V$ and at any iteration~$t\geq 1$, we write the posterior GP mean and variance given parameters~$a_{1:t}^\N{i}$, iterations $1,\ldots, t$, and the corresponding rewards~$y_{1:t}$ as~$\mu_t^\A{i}(\cdot)$ and~$\sigma_t^\A{i}(\cdot)$, respectively.
Moreover, we denote the GP mean and variance evaluated at~$a^\N{i}$ and time step~$\tilde t$ by~$\mu_t^\A{i}(a^\N{i},\tilde t)$ and~$\sigma_t^\A{i}(a^\N{i},\tilde t)$, respectively.
We can now use the GP mean and variance to estimate how the reward changes.
Specifically, we model the changes in the reward function caused by the neighboring agents by evaluating~$\mu_t^\A{i}$ and~$\sigma_t^\A{i}$ at different~$a^\N{i}$.
Further, we model the changes caused by non-neighboring agents in the time domain. 
\begin{mybox}
\tool{tool:time-series}{
To extrapolate the behavior of non-neighboring agents, we perform a one-step time-series prediction, \ie a one-step extrapolation in the time domain.
}
\end{mybox}
Tool~\ref{tool:time-series} enables the parameters of non-neighboring agents to be approximated implicitly by evaluating the GP mean~$\mu_t$ at time step~$\tilde t=t+1$.
This is a standard discrete time prediction setting with GPs, as described in, \eg\cite{roberts2013gaussian}.

\subsection{Spatio-temporal kernel}\label{sec:spatio-temporal}
The structure of the inputs of the GP mean~$\mu_t^\A{i}$ and variance~$\sigma_t^\A{i}$ induces a separation between the control parameters of the neighboring agents and time, \ie we have a \emph{spatio-temporal} separation. 
Therefore, it is natural to leverage a spatio-temporal kernel of the form
\begin{align}\label{eq:spatio-temporal}
    k((a^{(\mathcal{N}^{i}_+)}, \tilde t), (a^{(\mathcal{N}^{i}_+)'}, \tilde t^\prime))
    = k_\mathrm{S}(a^{(\mathcal{N}^{i}_+)}, a^{(\mathcal{N}^{i}_+)'}) k_\mathrm{T}(\tilde t, \tilde t^\prime)
\end{align}
to model the GP and thus the time-varying reward functions~$f_t^\A{i}$.
Besides our work, references on
time-varying BO~\cite{brunzema2022controller, bogunovic2016time, shao2024time} typically exploit a spatio-temporal kernel. 

\begin{mybox}
\tool{tool:spatio-temporal}{
We propose a custom spatio-temporal kernel that separately estimates the behavior of neighboring and non-neighboring agents to model the GP for each agent~$i\in V$ and each iteration~$t\geq 1$. 
}
\end{mybox}
Tool~\ref{tool:spatio-temporal} enables us to include prior knowledge into modeling the time-varying reward function~$f_t^\A{i}$ in both the spatial and temporal domains.

\subsubsection*{Spatial kernel} The spatial part~$k_\mathrm{S}$ of~\eqref{eq:spatio-temporal} encodes the covariance between the function values at two different parameters~$a^\N{i}, a^{(\mathcal{N}^i_+)'}$ at a fixed iteration step~$t$.
This is equivalent to the single-agent setting, where many safe BO algorithms rely on the smoothness of the underlying reward function~\cite{sui2015safe, berkenkamp2023bayesian}.
Therefore, smooth kernels such as the radial basis function (RBF) kernel or the Matérn kernel with a rather large~$\nu$ parameter are suitable~\cite{duvenaud2014automatic}.
We choose the Matérn52 kernel, \ie the Matérn kernel with~$\nu=\nicefrac{5}{2}$ as~$k_\mathrm{S}$.

\subsubsection*{Expected behavior in the time domain}

\begin{figure}
    \centering
    \input{Figures/final/agents_0_POV}
    \caption{\emph{Temporal change of the reward function.}
    Change of the rewards for a three-agent MAS without communication. 
    Agent~1 remains constant, while the other two agents optimize their control parameters, corresponding to a temporal change in the reward function from Agent~1's perspective.
    }
    \label{fig:time_series_POV}
\end{figure}
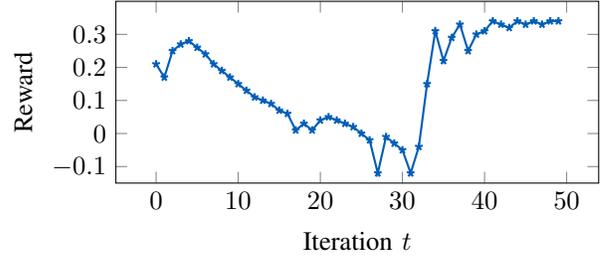
For the temporal part, we seek to encode the expected behavior of the reward function when the control parameters of non-neighboring agents change and the control parameters of neighboring agents remain constant.
To further investigate the temporal change of the reward function, we examine an example.
Figure~\ref{fig:time_series_POV} shows a reward trajectory in a three-agent MAS without communication over the number of iterations.
The parameters of the first agent remain constant, while the remaining two agents conduct safe BO to tune their control parameters.
Since the agents are isolated, the behavior in Figure~\ref{fig:time_series_POV} represents the temporal change in the reward function for this specific toy example, \ie how the reward changes when non-neighboring agents apply different control parameters.
Figure~\ref{fig:time_series_POV} shows that the reward value changes relatively smoothly at the beginning and towards the end of the optimization process.
Around the midpoint ($t\approx 25)$, there is a more abrupt change in the reward.
This behavior is expected since we are in a setting where each agent tunes its control parameters using safe BO. 
Initially, each agent is cautious, having a small set of parameters from which it can choose.
Towards the midpoint, the agents have a larger set of parameters that they believe to be safe and can therefore explore.
Approaching the end, the agents converge towards control parameters they believe to be optimal.
Let us now encode the behavior depicted in Figure~\ref{fig:time_series_POV} in a custom temporal kernel to assist the one-step time-series prediction~\ref{tool:time-series}.

\subsubsection*{Temporal kernel} To capture the smoothness in the beginning and towards the end of the optimization process, we want the temporal kernel to be dominated by smooth kernels. 
Therefore, we choose the RBF kernel~$k_\mathrm{RBF}$ that induces infinitely differentiable functions.
In contrast, the rougher parts can be represented by less smooth kernels.
Hence, we choose the Matérn12 kernel~$k_\mathrm{Ma12}$, which corresponds to the Ornstein-Uhlenbeck process with continuous but non-differentiable sample paths.
To achieve the desired properties, we combine the RBF and Matérn12 kernels.

\begin{figure}
\begin{center}
\begin{tikzpicture}[xscale=.725, yscale=.81]
\node[inner sep=0pt] (whitehead) at (4.155, 2.8)
     {\includegraphics[width=6cm]{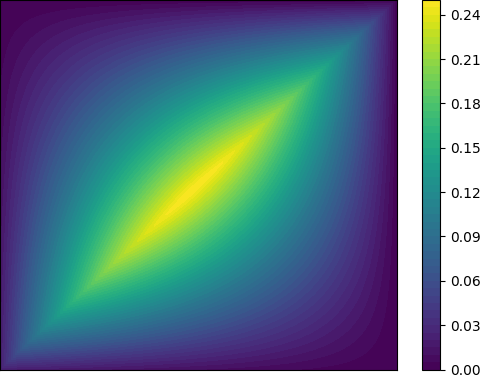}};
    \node[right, rotate=90, yshift=-1em, xshift=-2em] at (whitehead.east) { $k_\mathrm{W}(t,t^\prime)$}; 
\begin{axis}[
tick align=outside,
tick pos=left,
xlabel={$t$},
xmin=0, xmax=50,
xtick style={color=black},
ylabel={$t^\prime$},
ymin=0,
ymax=50,
ytick style={color=black}
]
\end{axis}
\end{tikzpicture}
\end{center}
    \caption{Weighting kernel~$k_\mathrm{W}(t,t^\prime)$ with~$T=50$.}
    \label{fig:weight}
\end{figure}

To combine both kernels, we create a third kernel~$k_\mathrm{W}$ that acts as the weighting kernel.
In particular, we choose  $k_\mathrm{W}:[1, T]^2\rightarrow \R_+$ as 
\begin{align}\label{eq:weight}
k_\mathrm{W}(t,t^\prime)=\frac{1}{T^2}\min(t,t^\prime)\cdot\min(T-t,T-t^\prime)
\end{align}
and obtain
\begin{align}\label{eq:temporal_kernel}
    k_\mathrm{T}(t,t^\prime)=k_\mathrm{RBF}(t,t^\prime) + k_\mathrm{W}(t,t^\prime)\cdot k_\mathrm{Ma12}(t,t^\prime)
\end{align}
as the resulting temporal kernel.
The kernel~$k_\mathrm{W}$~\eqref{eq:weight} is a product of the classic Brownian motion kernel and a ``reverse'' Brownian motion kernel, and is hence a valid kernel for all inputs~$(t,t^\prime) \in [0,T]^2.$
Importantly, we cannot work with a simple convex combination of~$k_\mathrm{RBF}$ and~$k_\mathrm{Ma12}$ as we need to preserve positive-definiteness to have a valid (reproducing) kernel~\cite[Section~4.1]{Steinwart2008SVM}.
This is ensured by the weighting kernel~$k_\mathrm{W}$ due to the fact that the products and sums of positive-definite kernels yield positive-definite kernels~\cite{aronszajn1950theory}.
Figure~\ref{fig:weight} illustrates our weighting kernel~$k_\mathrm{W}$.
As desired, the weighting of the Matérn12 kernel peaks around the midpoint~$\nicefrac{T}{2}$, introducing more abrupt changes.

\subsubsection*{Random functions using~$k_\mathrm{T}$}
\begin{figure}
    \centering
    \input{Figures/final/sample_paths}
    \caption{\emph{Random functions using~$k_\mathrm{T}$.}
    Two random functions created with the temporal kernel~$k_\mathrm{T}(t,t^\prime)$~\eqref{eq:temporal_kernel} with~$T=50$.
    For~$k_\mathrm{RBF}$, we use a lengthscale of~$\ell_\mathrm{RBF}=5$ and an output variance of~$\sigma_\mathrm{f, RBF}=1$, while~$k_\mathrm{Ma12}$ has~$\ell_{Ma12}=1$ and output variance~$\sigma_\mathrm{f, Ma12}=10$.
    }
    \label{fig:random_RKHS_temporal}
\end{figure}
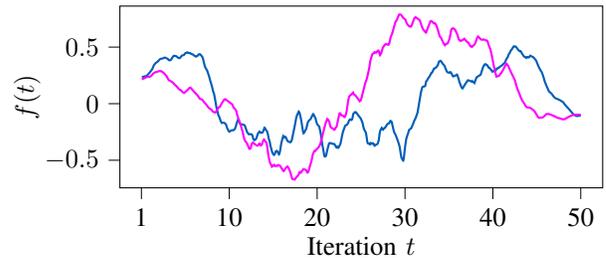

To assess whether the proposed temporal kernel~\eqref{eq:temporal_kernel} can capture behaviors like the one illustrated in Figure~\ref{fig:time_series_POV}, we create random functions using the kernel~\eqref{eq:temporal_kernel}. 
Specifically, we construct random functions that lie in the reproducing kernel Hilbert space (RKHS) of the kernel~$k_\mathrm{T}$ by using the pre-RKHS approach described in~\cite[Appendix~C.1]{Fiedler2021Practical}.
Figure~\ref{fig:random_RKHS_temporal} shows two random functions using kernel~$k_\mathrm{T}$, supporting the design choice of the proposed temporal kernel~\eqref{eq:temporal_kernel} as the functions exhibit similar smoothness properties as the toy example in Figure~\ref{fig:time_series_POV}.

\subsection{Safe Bayesian optimization}\label{sec:safe_BO}
Next, we describe our safe BO algorithm reminiscent of \safeopt~\cite{sui2015safe, berkenkamp2023bayesian} to solve~\eqref{eq:local_opt}.
As mentioned in Section~\ref{sec:time-varying}, solving~\eqref{eq:local_opt} serves as a proxy for solving the 
intractable optimization problem~\eqref{eq:global_opt}.

\subsubsection*{\safeopt} 
We use GP regression with the spatio-temporal kernel~\eqref{eq:spatio-temporal} to model the time-varying local reward functions~$f_t^\A{i}$ and solve the optimization problem~\eqref{eq:local_opt} by sequentially sampling the reward function~$f$.
To evaluate the reward function~$f$ only with parameters yielding rewards above the safety threshold~$h$, the  prediction uncertainty needs to be quantified. 
Hence,
we use the confidence intervals~$Q_t^{(i)}$ from~\cite{abbasi2013online} that bound the deviation between~$f_t^\A{i}$ and~$\mu_t^\A{i}$ with high probability.
To this end, we assume that~$f_t^{(i)}$ is a member of the RKHS of the spatio-temporal kernel~$k$~\eqref{eq:spatio-temporal} with known RKHS norm upper bound~$B\geq \|f_t^\A{i}\|_k$ for all agents~$i\in V$ and for all iterations~$t\geq 1$.
Then, by combining~\cite[Theorem~3.11]{abbasi2013online} with~\cite[Remark~3.13]{abbasi2013online}, the confidence intervals are 
$Q_t^\A{i}(\cdot) \coloneqq \mu_t^\A{i}(\cdot) \pm \beta_t^\A{i}\sigma_t^\A{i}(\cdot)$, where~$\beta_t^\A{i}$ is a data-dependent scalar that, among others, depends on the RKHS norm upper bound~$B$.
Moreover, we define the lower and upper confidence bounds as $\ell_t^\A{i}(\cdot) \coloneqq \min Q_t^\A{i}(\cdot)$ and $u_t^\A{i}(\cdot) \coloneqq \max Q_t^\A{i}(\cdot)$, respectively.
Then, we introduce the safe set~$S_t^\A{i}$ as the set of parameters for which the lower confidence bound~$\ell_t^\A{i}$ is larger than the safety threshold~$h$.
Different from~\cite{sui2015safe}, we define the safe set using the RKHS norm induced continuity from~\cite[Lemma~1]{tokmak2025automatic} instead of additionally requiring the Lipschitz constant.
We further  construct a set of potential maximizers~$M_t^\A{i}$ and potential expanders~$G_t^\A{i}$ to safely balance exploration and exploitation.
See~\cite{sui2015safe} for a detailed introduction to the definitions of the aforementioned sets.

\subsubsection*{Proposed algorithm}
Algorithm~\ref{alg:BO} summarizes our proposed safe BO algorithm.
To start the parameter optimization, we require a set of initial parameters~$\a_0$ that correspond to a safe experiment.
Further, we require the agents~$V$, the spatio-temporal kernel~$k$, the maximum number of iterations~$T$, the RKHS norm upper bound~$B$, the safety threshold~$h$, and access to the sampling oracle~$f$.

Then, each agent~$i$ computes the GP mean~$\mu_t^\A{i}$ and variance~$\sigma_t^\A{i}$ (\lline~3).
Notably, the GP mean and variance are evaluated for all possible control parameters of the neighboring agents~$a^\N{i}\in\domain^{\lvert \mathcal N^i_+\rvert }$.
Conversely, we evaluate~$\mu_t^\A{i}$ and~$\sigma_t^\A{i}$ only at~$t+1$ in the time domain.
Essentially, we are executing regression in the spatial domain and a one-step time-series prediction~\ref{tool:time-series}, \ie a one-step extrapolation in the time domain.

Next, every agent computes its confidence intervals (\lline~4), lower and upper confidence bounds (\lline~4), and its sets of safe parameters, potential maximizers, and potential expanders (\lline~5).
Then, each agent saves the most uncertain control parameter~$a_{t+1}^\N{i}$ that is either a potential maximizer or a potential expander (\lline~6) and projects~$a_{t+1}^\N{i}$ to its own parameter~$a_{t+1}^\A{i}$ (\lline~7).
As with other distributed methods~\cite{zerefa2024distributed, boyd2011distributed}, the computations for each agent~$i\in V$ in lines \lline~3-7 are parallelizable.

Furthermore, we introduce a sequential expert in every iteration, which is a protocol that is implementable a priori (\lline~8).
If agent~$j$ is the expert of round~$t$, each neighbor~$i\in\mathcal{N}(j)$ applies~$a_{t+1}^\N{j}[i]$ instead of~$a_{t+1}^\N{i}[i]$. 
That is, its prediction is overwritten by the expert~$j$ of iteration~$t$.
The sequential expert setting has been heuristically shown to improve exploration for two main reasons.
First, some agents may have a finer discretization due to \eg 
 more computing power or fewer neighboring agents.
A finer discretization density improves exploration of discretized safe BO algorithms~\cite[Section~3.3]{tokmak2024safe}.
Second, the sequential expert protocol helps to prevent the exploration from collapsing.\footnote{
Consider a setting in which agent~$i$ with~$\lvert \mathcal{N}_+^i\rvert=2$ computes~$a_{t+1}^\A{i}=[0,1]$ (\lline~7) and receives~$a_{t+1}^\A{j}=0$ from its neighbor~$j\in \mathcal{N}^j$ (\lline~9).
This yields the data point~$[0,0]\in a_{1:t}^\N{i}$ for agent~$i$.
Therefore, the parameter combination~$a_{t+1}^\A{i}=[0,1]$ remains uncertain and agent~$i$ may repeatedly suggest~$a_{t+1}^\A{i} = [0,1]$ without ever obtaining a corresponding sample.
The sequential expert setting mitigates this scenario.
}
Subsequently, the agents communicate their control parameters to their neighbors (\lline~9).
Then, we conduct an experiment (\lline~10) and update the sample set given the applied control parameters of the neighboring agents and the observed reward (\lline~11).
We repeat the procedure for~$T$ iterations and define the best control parameter as the one that corresponds to the highest reward value (\lline~12).

\begin{algorithm}
\caption{Safe BO for distributed MAS}
\label{alg:BO}
\begin{algorithmic}[1]
\Require{Safe initial parameter~$\a_0$, $V$, $k$, $T$, $B$, $h$, $f$}
\For{$t\in\{1,\ldots, T\}$}
\For{$i\in V$}
\State Compute~$\mu_t^\A{i}(a^\N{i},t+1)$, $\sigma_t^\A{i}(a^\N{i},t+1)$
\Statex \phantom{jhsdfjf} given~$a_{1:t}^\N{i}$,~$1,\ldots, t$, ~$y_{1:t}$
\State Determine~$Q_t^\A{i}(a^\N{i},t+1)$ and confidence
\Statex \phantom{jsksjdj} bounds $\ell_t(a^\N{i},t+1)$, $u_t(a^\N{i},t+1)$
\State Calculate sets~$S_t^\A{i}, G_t^\A{i}$, and~$M_t^\A{i}$
\State $a_{t+1}^\N{i} \gets \underset{a^\N{i}\in M_t^\A{i}\cup G_t^\A{i}}{\argmax} \sigma_t(a^\N{i}, t+1)$ 
\State $a_{t+1}^\A{i} \gets a_{t+1}^\N{i}[i]$ \Comment{Agent~$i$'s parameter}
\EndFor
\State $a_{t+1}^\A{i} \gets a_{t+1}^\N{j}[i]$, $j=[(t-1)\%N]+1, i\in \mathcal N(j)$
\State Communicate~$a_{t+1}^\A{i}$ to agent~$j$ if $j\in\mathcal{N}(i)$
\State $y_{t+1} = f(\a_{t+1})+\epsilon_{t+1}$ \Comment{Conduct experiment}
\State Update sample set~$a_{1:t}^\N{i}$ and rewards $y_{1+t}$
\EndFor
\State \textbf{Return} $\argmax_{\a_t}y_{1:t}$ \Comment{Highest reward parameter}
\end{algorithmic}
\end{algorithm}

%% file: Figures/final/agents_0_POV.tex
\begin{tikzpicture}
\begin{axis}[
    xlabel={Iteration~$t$},
    ymin=-0.15, ymax=0.4,
    xtick={0,10,20,30,40,50},
    ytick={-0.1, 0, 0.1, 0.2, 0.3},
    legend pos=outer north east,
    mark size=1.5pt,
    width=8cm,
    height=4cm,
    ylabel=Reward
]

\addplot[
    color=aaltoBlue,
    mark=star,
    thick
] coordinates {
    (0,0.21)(1,0.17)(2,0.25)(3,0.27)(4,0.28)(5,0.26)(6,0.24)(7,0.21)
    (8,0.19)(9,0.17)(10,0.15)(11,0.13)(12,0.11)(13,0.10)(14,0.09)(15,0.07)
    (16,0.06)(17,0.01)(18,0.03)(19,0.01)(20,0.04)(21,0.05)(22,0.04)(23,0.03)
    (24,0.02)(25,0.00)(26,-0.02)(27,-0.12)(28,-0.01)(29,-0.03)(30,-0.05)
    (31,-0.12)(32,-0.04)(33,0.15)(34,0.31)(35,0.22)(36,0.29)(37,0.33)
    (38,0.25)(39,0.30)(40,0.31)(41,0.34)(42,0.33)(43,0.32)(44,0.34)
    (45,0.33)(46,0.34)(47,0.33)(48,0.34)(49,0.34)
};

\end{axis}
\end{tikzpicture}

%% file: Figures/final/sample_paths.tex
\begin{tikzpicture}

\definecolor{darkgray176}{RGB}{176,176,176}
\definecolor{darkorange25512714}{RGB}{255,127,14}
\definecolor{steelblue31119180}{RGB}{31,119,180}

\begin{axis}[
tick align=outside,
tick pos=left,
x grid style={darkgray176},
xlabel={Iteration~$t$},
xmin=-48.95, xmax=1049.95,
xtick style={color=black},
xtick={0,200,400,600,800,1000},
xticklabels={1,10,20,30,40,50},
y grid style={darkgray176},
ylabel={$f(t)$},
ymin=-0.74374543428421, ymax=0.862578499317169,
ytick style={color=black},
width=8cm,
height=4cm,
]
\addplot [thick, aaltoBlue]
table {%
1 0.233689695596695
2 0.234672263264656
3 0.235988825559616
4 0.237752854824066
5 0.239447057247162
6 0.240880489349365
7 0.241464853286743
8 0.241964995861053
9 0.242342263460159
10 0.243465051054955
11 0.244888573884964
12 0.247602090239525
13 0.250420153141022
14 0.253715991973877
15 0.258333802223206
16 0.262950509786606
17 0.266877919435501
18 0.271042078733444
19 0.276059806346893
20 0.28106901049614
21 0.286139249801636
22 0.291283547878265
23 0.296509385108948
24 0.301823437213898
25 0.30804505944252
26 0.317462563514709
27 0.327050685882568
28 0.335973888635635
29 0.342781394720078
30 0.347942471504211
31 0.352638781070709
32 0.357111096382141
33 0.360321760177612
34 0.3632692694664
35 0.366092324256897
36 0.369150906801224
37 0.372427821159363
38 0.375938534736633
39 0.379700362682343
40 0.383710026741028
41 0.38797789812088
42 0.392142802476883
43 0.395445704460144
44 0.396809637546539
45 0.397118151187897
46 0.397663652896881
47 0.398450046777725
48 0.399479776620865
49 0.400754153728485
50 0.402278304100037
51 0.404050648212433
52 0.406078845262527
53 0.407592266798019
54 0.4065802693367
55 0.405817031860352
56 0.403864830732346
57 0.401639074087143
58 0.399604469537735
59 0.397768139839172
60 0.396082073450089
61 0.391879320144653
62 0.386359632015228
63 0.384121477603912
64 0.383210182189941
65 0.383421421051025
66 0.385187983512878
67 0.387133985757828
68 0.392126888036728
69 0.398036003112793
70 0.40419265627861
71 0.410624504089355
72 0.417350828647614
73 0.420062869787216
74 0.423061639070511
75 0.423306435346603
76 0.422268450260162
77 0.421478450298309
78 0.421216815710068
79 0.420403361320496
80 0.416984260082245
81 0.413753718137741
82 0.410764783620834
83 0.407994449138641
84 0.405454218387604
85 0.403406858444214
86 0.402072459459305
87 0.403239220380783
88 0.40550422668457
89 0.409821629524231
90 0.414414763450623
91 0.419311702251434
92 0.424450874328613
93 0.42594912648201
94 0.427522957324982
95 0.428443938493729
96 0.429727643728256
97 0.431342005729675
98 0.435978621244431
99 0.441037476062775
100 0.446483492851257
101 0.448247998952866
102 0.449446469545364
103 0.451182842254639
104 0.453803211450577
105 0.45216965675354
106 0.450340330600739
107 0.448926776647568
108 0.44796034693718
109 0.447432547807693
110 0.447381854057312
111 0.447289794683456
112 0.443914741277695
113 0.440811693668365
114 0.438560992479324
115 0.437539607286453
116 0.43695256114006
117 0.43531346321106
118 0.431730389595032
119 0.428579151630402
120 0.425853908061981
121 0.423601299524307
122 0.421284139156342
123 0.417919158935547
124 0.416195034980774
125 0.418342590332031
126 0.419756323099136
127 0.421622514724731
128 0.424007594585419
129 0.429171204566956
130 0.434880137443542
131 0.436737596988678
132 0.437040328979492
133 0.437773883342743
134 0.434620946645737
135 0.431180983781815
136 0.427299797534943
137 0.421494156122208
138 0.416247725486755
139 0.411598563194275
140 0.400665104389191
141 0.384329617023468
142 0.368219196796417
143 0.352503895759583
144 0.337169587612152
145 0.323725432157516
146 0.316727995872498
147 0.309984296560287
148 0.303588390350342
149 0.29751980304718
150 0.289527416229248
151 0.278359979391098
152 0.266117870807648
153 0.254128664731979
154 0.242423117160797
155 0.228473037481308
156 0.211994960904121
157 0.199728921055794
158 0.187562242150307
159 0.174641475081444
160 0.160907447338104
161 0.147231578826904
162 0.130359560251236
163 0.112539172172546
164 0.101209223270416
165 0.0916608572006226
166 0.0749487280845642
167 0.0535002499818802
168 0.0318953990936279
169 0.0100322663784027
170 -0.012151226401329
171 -0.0348343849182129
172 -0.0520360022783279
173 -0.0669239684939384
174 -0.0821616351604462
175 -0.0979463160037994
176 -0.108620844781399
177 -0.114354722201824
178 -0.125152349472046
179 -0.136487036943436
180 -0.148383721709251
181 -0.157492190599442
182 -0.163777604699135
183 -0.170765906572342
184 -0.169504463672638
185 -0.168659463524818
186 -0.168385341763496
187 -0.168636858463287
188 -0.169407874345779
189 -0.170700132846832
190 -0.17294754087925
191 -0.177078694105148
192 -0.181738823652267
193 -0.188200160861015
194 -0.200456634163857
195 -0.213263928890228
196 -0.222735047340393
197 -0.229695811867714
198 -0.237290143966675
199 -0.244192436337471
200 -0.247885778546333
201 -0.245887503027916
202 -0.243126034736633
203 -0.240357935428619
204 -0.238135427236557
205 -0.236445277929306
206 -0.235272824764252
207 -0.234619185328484
208 -0.234456092119217
209 -0.234788909554482
210 -0.233415871858597
211 -0.229321151971817
212 -0.225933089852333
213 -0.221704423427582
214 -0.211137473583221
215 -0.198294430971146
216 -0.185729756951332
217 -0.173429876565933
218 -0.161342784762383
219 -0.149440541863441
220 -0.141147270798683
221 -0.143172904849052
222 -0.152896165847778
223 -0.162902966141701
224 -0.173120737075806
225 -0.179104700684547
226 -0.182355850934982
227 -0.182486727833748
228 -0.182801276445389
229 -0.183292388916016
230 -0.183968439698219
231 -0.184817314147949
232 -0.185842871665955
233 -0.187037914991379
234 -0.188405498862267
235 -0.191147074103355
236 -0.201295673847198
237 -0.211761459708214
238 -0.222450852394104
239 -0.233390048146248
240 -0.244592890143394
241 -0.252842783927917
242 -0.258125185966492
243 -0.263711363077164
244 -0.26961675286293
245 -0.270325481891632
246 -0.269741088151932
247 -0.269452661275864
248 -0.26945772767067
249 -0.269736915826797
250 -0.270317435264587
251 -0.271152049303055
252 -0.27737095952034
253 -0.284929752349854
254 -0.300567150115967
255 -0.313859790563583
256 -0.319143384695053
257 -0.323223352432251
258 -0.321540415287018
259 -0.31696754693985
260 -0.315102458000183
261 -0.310041010379791
262 -0.303116053342819
263 -0.290177881717682
264 -0.278267621994019
265 -0.266140073537827
266 -0.258779883384705
267 -0.256035357713699
268 -0.251465737819672
269 -0.246255904436111
270 -0.250632435083389
271 -0.256367444992065
272 -0.257299572229385
273 -0.256947427988052
274 -0.256771951913834
275 -0.256689071655273
276 -0.251992791891098
277 -0.245568335056305
278 -0.235892772674561
279 -0.224164217710495
280 -0.212391376495361
281 -0.213586956262589
282 -0.221229940652847
283 -0.228884726762772
284 -0.234955072402954
285 -0.240020513534546
286 -0.245606318116188
287 -0.256999343633652
288 -0.271599471569061
289 -0.291979849338531
290 -0.31187704205513
291 -0.323062539100647
292 -0.334461688995361
293 -0.355579227209091
294 -0.378216981887817
295 -0.389398694038391
296 -0.405018150806427
297 -0.418349474668503
298 -0.432129859924316
299 -0.438945591449738
300 -0.444896638393402
301 -0.451432824134827
302 -0.453424453735352
303 -0.448055058717728
304 -0.441064357757568
305 -0.430263221263885
306 -0.420436024665833
307 -0.418329060077667
308 -0.416720598936081
309 -0.420405387878418
310 -0.426522314548492
311 -0.433173060417175
312 -0.440383672714233
313 -0.448028594255447
314 -0.444658279418945
315 -0.427652686834335
316 -0.408672362565994
317 -0.383350551128387
318 -0.358349442481995
319 -0.340094566345215
320 -0.324124336242676
321 -0.311049997806549
322 -0.296300053596497
323 -0.286113679409027
324 -0.291804075241089
325 -0.29712301492691
326 -0.295663237571716
327 -0.294338881969452
328 -0.293179333209991
329 -0.299524307250977
330 -0.308027744293213
331 -0.316764712333679
332 -0.325768738985062
333 -0.330731689929962
334 -0.334362506866455
335 -0.33643651008606
336 -0.337344288825989
337 -0.338551133871078
338 -0.340063482522964
339 -0.341882348060608
340 -0.338995933532715
341 -0.333390325307846
342 -0.324907958507538
343 -0.311385691165924
344 -0.292594641447067
345 -0.271169275045395
346 -0.253041565418243
347 -0.234604313969612
348 -0.2161925137043
349 -0.19735561311245
350 -0.177368476986885
351 -0.155393064022064
352 -0.132417619228363
353 -0.116138696670532
354 -0.104581207036972
355 -0.0960239171981812
356 -0.0871506333351135
357 -0.0779275596141815
358 -0.0697965621948242
359 -0.0741850435733795
360 -0.0819922089576721
361 -0.0969368517398834
362 -0.114845290780067
363 -0.132515162229538
364 -0.151339530944824
365 -0.176882088184357
366 -0.202332049608231
367 -0.227499678730965
368 -0.2482750415802
369 -0.261099755764008
370 -0.25785356760025
371 -0.250835657119751
372 -0.24391858279705
373 -0.237085968255997
374 -0.230322912335396
375 -0.223615169525146
376 -0.210670232772827
377 -0.192735195159912
378 -0.174755930900574
379 -0.16406461596489
380 -0.157782673835754
381 -0.1513611972332
382 -0.146728798747063
383 -0.143343105912209
384 -0.139790937304497
385 -0.13607157766819
386 -0.128879114985466
387 -0.120672717690468
388 -0.112787783145905
389 -0.10466256737709
390 -0.0969942510128021
391 -0.0920579135417938
392 -0.0962989032268524
393 -0.104089230298996
394 -0.111631855368614
395 -0.118947833776474
396 -0.12605445086956
397 -0.139801114797592
398 -0.158118396997452
399 -0.169064939022064
400 -0.179966121912003
401 -0.185876131057739
402 -0.190081924200058
403 -0.194265440106392
404 -0.198443979024887
405 -0.204902976751328
406 -0.225410223007202
407 -0.241765946149826
408 -0.259836256504059
409 -0.287009835243225
410 -0.31442803144455
411 -0.342171996831894
412 -0.365947544574738
413 -0.386990338563919
414 -0.408542513847351
415 -0.430660843849182
416 -0.453398793935776
417 -0.466188579797745
418 -0.466225981712341
419 -0.466351389884949
420 -0.460787862539291
421 -0.448909521102905
422 -0.430726379156113
423 -0.413127809762955
424 -0.396127641201019
425 -0.380947679281235
426 -0.367345124483109
427 -0.354921340942383
428 -0.348554372787476
429 -0.348190486431122
430 -0.348231673240662
431 -0.348675161600113
432 -0.349524289369583
433 -0.350780367851257
434 -0.354137092828751
435 -0.355660378932953
436 -0.355197370052338
437 -0.355156600475311
438 -0.355539411306381
439 -0.356345415115356
440 -0.361119151115417
441 -0.369533628225327
442 -0.375731468200684
443 -0.381166517734528
444 -0.387097746133804
445 -0.383971691131592
446 -0.380457758903503
447 -0.381982415914536
448 -0.389582872390747
449 -0.386568486690521
450 -0.378172188997269
451 -0.370260179042816
452 -0.362811088562012
453 -0.355708956718445
454 -0.33668926358223
455 -0.317268341779709
456 -0.298169016838074
457 -0.279343008995056
458 -0.260740429162979
459 -0.246976971626282
460 -0.235843569040298
461 -0.224823743104935
462 -0.213888972997665
463 -0.203010767698288
464 -0.194276332855225
465 -0.192109376192093
466 -0.18994414806366
467 -0.187775850296021
468 -0.185598373413086
469 -0.181619167327881
470 -0.173273980617523
471 -0.168413281440735
472 -0.154154270887375
473 -0.147233694791794
474 -0.143030405044556
475 -0.138680100440979
476 -0.139912873506546
477 -0.142783671617508
478 -0.144420683383942
479 -0.140794187784195
480 -0.134756326675415
481 -0.114803612232208
482 -0.0948950350284576
483 -0.0853883624076843
484 -0.0829394459724426
485 -0.0808640122413635
486 -0.0747276544570923
487 -0.0748788416385651
488 -0.074733167886734
489 -0.0783199369907379
490 -0.0837922096252441
491 -0.0889919102191925
492 -0.0939333140850067
493 -0.0989276170730591
494 -0.104471474885941
495 -0.109798312187195
496 -0.114922642707825
497 -0.119858592748642
498 -0.124619275331497
499 -0.129217833280563
500 -0.13732984662056
501 -0.146615564823151
502 -0.155798211693764
503 -0.164902061223984
504 -0.181850850582123
505 -0.19985918700695
506 -0.217904701828957
507 -0.227404847741127
508 -0.234597980976105
509 -0.241921916604042
510 -0.249397158622742
511 -0.257029056549072
512 -0.278128415346146
513 -0.305544912815094
514 -0.333277344703674
515 -0.348365038633347
516 -0.351899862289429
517 -0.353191256523132
518 -0.354933023452759
519 -0.357130944728851
520 -0.359792530536652
521 -0.351772427558899
522 -0.347756087779999
523 -0.352388918399811
524 -0.354471296072006
525 -0.355325162410736
526 -0.354199945926666
527 -0.349654406309128
528 -0.352316319942474
529 -0.358668833971024
530 -0.367798209190369
531 -0.371151983737946
532 -0.365367233753204
533 -0.364073008298874
534 -0.363514304161072
535 -0.361660122871399
536 -0.358971446752548
537 -0.357060670852661
538 -0.351406574249268
539 -0.33835232257843
540 -0.325755536556244
541 -0.302801311016083
542 -0.26985114812851
543 -0.247948855161667
544 -0.235134482383728
545 -0.227922886610031
546 -0.220893383026123
547 -0.214194551110268
548 -0.21915178000927
549 -0.226516336202621
550 -0.2299844622612
551 -0.225388094782829
552 -0.220983609557152
553 -0.216761916875839
554 -0.213104069232941
555 -0.21143838763237
556 -0.204454362392426
557 -0.19761648774147
558 -0.191546410322189
559 -0.185601875185966
560 -0.179774045944214
561 -0.177669614553452
562 -0.17601178586483
563 -0.17444896697998
564 -0.172978401184082
565 -0.163855135440826
566 -0.154533594846725
567 -0.151686176657677
568 -0.154608696699142
569 -0.157588064670563
570 -0.164861559867859
571 -0.182632640004158
572 -0.200695663690567
573 -0.217026039958
574 -0.233556866645813
575 -0.24899435043335
576 -0.255783528089523
577 -0.263837337493896
578 -0.272265523672104
579 -0.281094878911972
580 -0.290352910757065
581 -0.300065964460373
582 -0.310264408588409
583 -0.320979714393616
584 -0.332241654396057
585 -0.344084978103638
586 -0.356545299291611
587 -0.369657605886459
588 -0.384247809648514
589 -0.409830152988434
590 -0.436183393001556
591 -0.459632754325867
592 -0.479536116123199
593 -0.486062794923782
594 -0.492649346590042
595 -0.500277757644653
596 -0.502881646156311
597 -0.488550961017609
598 -0.459581136703491
599 -0.420336067676544
600 -0.386774122714996
601 -0.358338624238968
602 -0.332091957330704
603 -0.308829009532928
604 -0.296083629131317
605 -0.283942192792892
606 -0.272568941116333
607 -0.263563126325607
608 -0.248546749353409
609 -0.234032437205315
610 -0.230266571044922
611 -0.226531505584717
612 -0.22325000166893
613 -0.220420598983765
614 -0.218072861433029
615 -0.208643734455109
616 -0.199224427342415
617 -0.190214425325394
618 -0.181620761752129
619 -0.173174202442169
620 -0.160040080547333
621 -0.151729628443718
622 -0.143738925457001
623 -0.136055141687393
624 -0.128676488995552
625 -0.117731541395187
626 -0.106792345643044
627 -0.0960840210318565
628 -0.0855894684791565
629 -0.0684489607810974
630 -0.0451422780752182
631 -0.0219402015209198
632 0.00120286643505096
633 0.024344727396965
634 0.0475395619869232
635 0.0708418488502502
636 0.0943378210067749
637 0.11047637462616
638 0.127161800861359
639 0.142127349972725
640 0.147305652499199
641 0.150659739971161
642 0.150903150439262
643 0.151503637433052
644 0.152454659342766
645 0.153751283884048
646 0.157278165221214
647 0.175362303853035
648 0.195089414715767
649 0.200954213738441
650 0.203540444374084
651 0.211822405457497
652 0.220540165901184
653 0.22527651488781
654 0.23048323392868
655 0.236175253987312
656 0.242359697818756
657 0.249049142003059
658 0.256256252527237
659 0.263986229896545
660 0.272369235754013
661 0.273382306098938
662 0.278088957071304
663 0.285782605409622
664 0.294051647186279
665 0.302913844585419
666 0.310446381568909
667 0.310995995998383
668 0.312875300645828
669 0.315387606620789
670 0.318520933389664
671 0.322279453277588
672 0.326658874750137
673 0.330869376659393
674 0.336378812789917
675 0.342974543571472
676 0.356800228357315
677 0.368986129760742
678 0.37375482916832
679 0.375167191028595
680 0.37728488445282
681 0.378874093294144
682 0.37427344918251
683 0.37000098824501
684 0.363585650920868
685 0.357848346233368
686 0.352763652801514
687 0.348315834999084
688 0.344487190246582
689 0.339760541915894
690 0.334546089172363
691 0.324745416641235
692 0.314070165157318
693 0.303904473781586
694 0.294876396656036
695 0.284875631332397
696 0.279892981052399
697 0.277749180793762
698 0.275983840227127
699 0.274580717086792
700 0.27353847026825
701 0.274976670742035
702 0.277054339647293
703 0.279472649097443
704 0.277618050575256
705 0.271331071853638
706 0.265390038490295
707 0.258674621582031
708 0.248900577425957
709 0.242307141423225
710 0.240879118442535
711 0.24223604798317
712 0.243780687451363
713 0.24552121758461
714 0.2474544942379
715 0.249577224254608
716 0.251890003681183
717 0.254414856433868
718 0.257057547569275
719 0.252781063318253
720 0.242566227912903
721 0.230934605002403
722 0.219465374946594
723 0.204517051577568
724 0.187693014740944
725 0.179688081145287
726 0.171644076704979
727 0.159452497959137
728 0.149747908115387
729 0.140977382659912
730 0.137515008449554
731 0.134894907474518
732 0.140629947185516
733 0.147366881370544
734 0.153906136751175
735 0.160288944840431
736 0.166510254144669
737 0.172592505812645
738 0.178532347083092
739 0.184356704354286
740 0.190066307783127
741 0.195673018693924
742 0.201186746358871
743 0.206640169024467
744 0.206944584846497
745 0.205624178051949
746 0.204225823283195
747 0.202751219272614
748 0.20118573307991
749 0.199539214372635
750 0.197784617543221
751 0.195948243141174
752 0.197510421276093
753 0.199986845254898
754 0.196262940764427
755 0.192424193024635
756 0.188438981771469
757 0.184314876794815
758 0.181803569197655
759 0.192863836884499
760 0.203554213047028
761 0.21408349275589
762 0.224447950720787
763 0.2330062687397
764 0.235731065273285
765 0.238371968269348
766 0.240944504737854
767 0.243436187505722
768 0.245861172676086
769 0.248291611671448
770 0.250655025243759
771 0.252927958965302
772 0.255322456359863
773 0.266351997852325
774 0.278629541397095
775 0.290822356939316
776 0.303721606731415
777 0.314913213253021
778 0.32613730430603
779 0.334266245365143
780 0.339061230421066
781 0.344011038541794
782 0.349097013473511
783 0.349712640047073
784 0.347236424684525
785 0.344916582107544
786 0.342742115259171
787 0.340846478939056
788 0.334289342164993
789 0.327594429254532
790 0.320983231067657
791 0.316988855600357
792 0.317511409521103
793 0.312571167945862
794 0.307604670524597
795 0.30267345905304
796 0.298622906208038
797 0.295048713684082
798 0.288588494062424
799 0.282579600811005
800 0.283987939357758
801 0.285178661346436
802 0.286304861307144
803 0.286992937326431
804 0.286000728607178
805 0.284895181655884
806 0.288490951061249
807 0.296893894672394
808 0.305240213871002
809 0.311571151018143
810 0.312417328357697
811 0.312276303768158
812 0.312084555625916
813 0.315847337245941
814 0.319437056779861
815 0.323018342256546
816 0.326600551605225
817 0.329359799623489
818 0.331603348255157
819 0.333868265151978
820 0.336164832115173
821 0.339617043733597
822 0.343310207128525
823 0.34704378247261
824 0.350849241018295
825 0.354711979627609
826 0.358647853136063
827 0.362659603357315
828 0.368964046239853
829 0.378946512937546
830 0.389094978570938
831 0.394958555698395
832 0.400379598140717
833 0.406044393777847
834 0.414272785186768
835 0.426251530647278
836 0.435017108917236
837 0.442598760128021
838 0.446067631244659
839 0.454536229372025
840 0.463339567184448
841 0.469678848981857
842 0.47499293088913
843 0.479848921298981
844 0.48562628030777
845 0.491847306489944
846 0.498531937599182
847 0.503811717033386
848 0.507428407669067
849 0.506693005561829
850 0.50633829832077
851 0.504256308078766
852 0.504196286201477
853 0.500267565250397
854 0.495984733104706
855 0.489867150783539
856 0.484284967184067
857 0.479233980178833
858 0.474696904420853
859 0.470683038234711
860 0.467155396938324
861 0.464103937149048
862 0.4654501080513
863 0.464728087186813
864 0.459725320339203
865 0.452294915914536
866 0.445425391197205
867 0.439021319150925
868 0.433080017566681
869 0.42758446931839
870 0.420552492141724
871 0.417014062404633
872 0.414870202541351
873 0.413094222545624
874 0.409794509410858
875 0.405658721923828
876 0.404036730527878
877 0.407856643199921
878 0.4122414290905
879 0.412133753299713
880 0.409820646047592
881 0.408847242593765
882 0.409367173910141
883 0.413284838199615
884 0.419221878051758
885 0.424376368522644
886 0.422600865364075
887 0.420192003250122
888 0.418314307928085
889 0.416655123233795
890 0.414179235696793
891 0.412222683429718
892 0.410806983709335
893 0.407209575176239
894 0.401858359575272
895 0.397040903568268
896 0.392758905887604
897 0.389028251171112
898 0.382522284984589
899 0.376085668802261
900 0.366575241088867
901 0.356369853019714
902 0.346694231033325
903 0.337509989738464
904 0.328805029392242
905 0.320549309253693
906 0.312727570533752
907 0.305331587791443
908 0.298336923122406
909 0.291724145412445
910 0.28550773859024
911 0.279635548591614
912 0.270479261875153
913 0.260773152112961
914 0.249089032411575
915 0.236984670162201
916 0.223755478858948
917 0.210244983434677
918 0.197236195206642
919 0.186948210000992
920 0.176907181739807
921 0.167682290077209
922 0.157654047012329
923 0.148774683475494
924 0.1404048204422
925 0.132150709629059
926 0.122931659221649
927 0.113637149333954
928 0.107459850609303
929 0.10203392803669
930 0.100533492863178
931 0.0980255454778671
932 0.0941891744732857
933 0.0903226509690285
934 0.0858361199498177
935 0.0808163732290268
936 0.0758106037974358
937 0.07050921022892
938 0.0646796450018883
939 0.0601436942815781
940 0.0561521425843239
941 0.0521257519721985
942 0.0491458773612976
943 0.0476294606924057
944 0.046041876077652
945 0.0444187670946121
946 0.0427667275071144
947 0.0410880222916603
948 0.039392463862896
949 0.0376800894737244
950 0.0359642654657364
951 0.0342474579811096
952 0.0325434133410454
953 0.0301034972071648
954 0.0275222212076187
955 0.0249774903059006
956 0.0230681523680687
957 0.0204326510429382
958 0.0183314681053162
959 0.0160297155380249
960 0.0127382576465607
961 0.00784417241811752
962 0.00162409991025925
963 -0.0034201517701149
964 -0.00815939903259277
965 -0.0127846300601959
966 -0.0173395052552223
967 -0.0218324065208435
968 -0.0262690857052803
969 -0.0306574255228043
970 -0.0350129157304764
971 -0.0389577746391296
972 -0.0430562049150467
973 -0.0472292602062225
974 -0.0513913854956627
975 -0.0560492426156998
976 -0.0606396794319153
977 -0.065201573073864
978 -0.0697445720434189
979 -0.0742444396018982
980 -0.0795559883117676
981 -0.0844125002622604
982 -0.0885366648435593
983 -0.0926734730601311
984 -0.0955018699169159
985 -0.0980590879917145
986 -0.100487388670444
987 -0.10301461815834
988 -0.105296894907951
989 -0.107423931360245
990 -0.108469322323799
991 -0.109250962734222
992 -0.109192505478859
993 -0.108364380896091
994 -0.107626631855965
995 -0.107104077935219
996 -0.106780506670475
997 -0.106640160083771
998 -0.106372274458408
999 -0.106279321014881
1000 -0.106345728039742
};
\addplot [thick, magenta]
table {%
1 0.215191662311554
2 0.216801688075066
3 0.218482241034508
4 0.220236718654633
5 0.222071200609207
6 0.223997473716736
7 0.226014852523804
8 0.228140205144882
9 0.230378821492195
10 0.231840550899506
11 0.233396172523499
12 0.235050544142723
13 0.236040830612183
14 0.236236676573753
15 0.236592918634415
16 0.237079337239265
17 0.236918777227402
18 0.236900895833969
19 0.237094268202782
20 0.23954838514328
21 0.242852672934532
22 0.245758026838303
23 0.248823642730713
24 0.252109616994858
25 0.257214576005936
26 0.262368500232697
27 0.265808790922165
28 0.26937460899353
29 0.271638929843903
30 0.274075865745544
31 0.276724457740784
32 0.27813732624054
33 0.278864204883575
34 0.279652744531631
35 0.2806436419487
36 0.281875252723694
37 0.283172279596329
38 0.284635156393051
39 0.286371141672134
40 0.288057625293732
41 0.288543671369553
42 0.288254618644714
43 0.286822319030762
44 0.287980943918228
45 0.286632031202316
46 0.284700632095337
47 0.283004999160767
48 0.280833661556244
49 0.277116030454636
50 0.273606806993484
51 0.270284205675125
52 0.267133712768555
53 0.262829542160034
54 0.258340179920197
55 0.253987312316895
56 0.249773025512695
57 0.244958192110062
58 0.237884253263474
59 0.23082435131073
60 0.223792776465416
61 0.216776415705681
62 0.209724172949791
63 0.206435829401016
64 0.204761624336243
65 0.202830076217651
66 0.200591266155243
67 0.198369964957237
68 0.196159243583679
69 0.19370986521244
70 0.191208451986313
71 0.188682168722153
72 0.183871418237686
73 0.179916560649872
74 0.1758693754673
75 0.171770647168159
76 0.167640700936317
77 0.163972899317741
78 0.161075919866562
79 0.157202005386353
80 0.153267651796341
81 0.149224281311035
82 0.144815742969513
83 0.140718519687653
84 0.135348081588745
85 0.130868494510651
86 0.12696860730648
87 0.124037072062492
88 0.121707245707512
89 0.120499178767204
90 0.119215399026871
91 0.114940449595451
92 0.109383843839169
93 0.105328239500523
94 0.102319851517677
95 0.0991488546133041
96 0.0956936851143837
97 0.0941970646381378
98 0.0976088345050812
99 0.101811751723289
100 0.103989109396935
101 0.106798827648163
102 0.110650934278965
103 0.114468559622765
104 0.118267826735973
105 0.122063539922237
106 0.125873774290085
107 0.129709869623184
108 0.133595705032349
109 0.137575298547745
110 0.141287088394165
111 0.140950426459312
112 0.140652745962143
113 0.136116877198219
114 0.13167043030262
115 0.127245053648949
116 0.122856169939041
117 0.118495553731918
118 0.114158943295479
119 0.109847500920296
120 0.105537995696068
121 0.101233080029488
122 0.0971864610910416
123 0.0934725478291512
124 0.0892734676599503
125 0.0850643292069435
126 0.0808363109827042
127 0.0765838697552681
128 0.0722909420728683
129 0.0668132752180099
130 0.0608161240816116
131 0.0547489151358604
132 0.0486194118857384
133 0.0435175150632858
134 0.0418257713317871
135 0.0406388342380524
136 0.0393899753689766
137 0.0380728393793106
138 0.0367218926548958
139 0.0326582342386246
140 0.0272597149014473
141 0.0217767134308815
142 0.0162166133522987
143 0.0119128450751305
144 0.0076722651720047
145 0.00331816077232361
146 -0.00115247815847397
147 -0.00574595481157303
148 -0.0104632899165154
149 -0.0153166130185127
150 -0.0203075632452965
151 -0.025452233850956
152 -0.0307487025856972
153 -0.0362174659967422
154 -0.0404022112488747
155 -0.0422735586762428
156 -0.0443648993968964
157 -0.0466093868017197
158 -0.0490236282348633
159 -0.0515972301363945
160 -0.0554118528962135
161 -0.0631063356995583
162 -0.0686835050582886
163 -0.0707370266318321
164 -0.0723297968506813
165 -0.0740104541182518
166 -0.0758573710918427
167 -0.0778955519199371
168 -0.0785696804523468
169 -0.0756243318319321
170 -0.07121592015028
171 -0.0635562762618065
172 -0.0559437647461891
173 -0.048392653465271
174 -0.040822334587574
175 -0.0337522700428963
176 -0.0331160798668861
177 -0.0325353443622589
178 -0.0301345661282539
179 -0.0240242034196854
180 -0.0175093933939934
181 -0.0106214880943298
182 -0.0032811164855957
183 0.00423739850521088
184 0.0100523680448532
185 0.0157770067453384
186 0.0217418521642685
187 0.0279628038406372
188 0.0344919115304947
189 0.0376154035329819
190 0.0392638593912125
191 0.040364533662796
192 0.0366633981466293
193 0.0333652645349503
194 0.0303802937269211
195 0.0284850299358368
196 0.0256021171808243
197 0.0231199711561203
198 0.0197691172361374
199 0.0161477625370026
200 0.012860506772995
201 0.0098845362663269
202 0.00722166895866394
203 0.0048927366733551
204 0.00256209075450897
205 -0.000956028699874878
206 -0.00418949127197266
207 -0.00762392580509186
208 -0.0166604518890381
209 -0.0270002633333206
210 -0.035370945930481
211 -0.0391155779361725
212 -0.0425957441329956
213 -0.0458528995513916
214 -0.0513471364974976
215 -0.0588217526674271
216 -0.065842479467392
217 -0.0817775726318359
218 -0.0993947684764862
219 -0.111395865678787
220 -0.122496962547302
221 -0.133516013622284
222 -0.14447557926178
223 -0.155414551496506
224 -0.166340798139572
225 -0.177281141281128
226 -0.18870609998703
227 -0.202783152461052
228 -0.217014253139496
229 -0.229838252067566
230 -0.241303190588951
231 -0.257650017738342
232 -0.276226133108139
233 -0.295055598020554
234 -0.310129970312119
235 -0.320400804281235
236 -0.329148948192596
237 -0.333794236183167
238 -0.337096065282822
239 -0.343240767717361
240 -0.349759101867676
241 -0.35945588350296
242 -0.364856958389282
243 -0.374960064888
244 -0.385456740856171
245 -0.396153092384338
246 -0.40009480714798
247 -0.39892590045929
248 -0.396753400564194
249 -0.387022197246552
250 -0.368993759155273
251 -0.358626902103424
252 -0.349785298109055
253 -0.343785554170609
254 -0.344669044017792
255 -0.34758448600769
256 -0.350590378046036
257 -0.353860437870026
258 -0.357307910919189
259 -0.360647559165955
260 -0.362827241420746
261 -0.365264266729355
262 -0.367977678775787
263 -0.370971351861954
264 -0.369925826787949
265 -0.368412017822266
266 -0.368465155363083
267 -0.361617386341095
268 -0.353451371192932
269 -0.35193943977356
270 -0.357130408287048
271 -0.362894922494888
272 -0.368879765272141
273 -0.366622686386108
274 -0.361117720603943
275 -0.353688895702362
276 -0.340963572263718
277 -0.328163623809814
278 -0.32300215959549
279 -0.316604971885681
280 -0.320680230855942
281 -0.326515018939972
282 -0.332414895296097
283 -0.338361829519272
284 -0.342357307672501
285 -0.352020025253296
286 -0.371537625789642
287 -0.399491906166077
288 -0.430414974689484
289 -0.456614315509796
290 -0.475236505270004
291 -0.494087845087051
292 -0.513378500938416
293 -0.525318384170532
294 -0.533280313014984
295 -0.537582516670227
296 -0.545945286750793
297 -0.557193219661713
298 -0.554721355438232
299 -0.550116062164307
300 -0.546053171157837
301 -0.542525291442871
302 -0.540201723575592
303 -0.545385897159576
304 -0.552173733711243
305 -0.552318930625916
306 -0.550782561302185
307 -0.549781918525696
308 -0.549299299716949
309 -0.549358367919922
310 -0.549426972866058
311 -0.542170643806458
312 -0.542065620422363
313 -0.543511271476746
314 -0.545453906059265
315 -0.547903418540955
316 -0.550863444805145
317 -0.556927621364594
318 -0.56419974565506
319 -0.569977760314941
320 -0.575819432735443
321 -0.582242846488953
322 -0.589274227619171
323 -0.594694495201111
324 -0.595732867717743
325 -0.59025377035141
326 -0.584128797054291
327 -0.578584611415863
328 -0.573056697845459
329 -0.567840039730072
330 -0.563164532184601
331 -0.559009790420532
332 -0.555380523204803
333 -0.55604487657547
334 -0.562696754932404
335 -0.569890022277832
336 -0.577634572982788
337 -0.586029052734375
338 -0.59634405374527
339 -0.607301831245422
340 -0.619333207607269
341 -0.634226322174072
342 -0.649904131889343
343 -0.65799742937088
344 -0.666768550872803
345 -0.667412281036377
346 -0.667398452758789
347 -0.667696177959442
348 -0.668807506561279
349 -0.670730710029602
350 -0.665210127830505
351 -0.659302830696106
352 -0.654187023639679
353 -0.649576961994171
354 -0.643794476985931
355 -0.64026927947998
356 -0.636942148208618
357 -0.63077700138092
358 -0.624264240264893
359 -0.614209294319153
360 -0.60512912273407
361 -0.604249179363251
362 -0.610941052436829
363 -0.609701454639435
364 -0.604787886142731
365 -0.599402904510498
366 -0.593095779418945
367 -0.587258338928223
368 -0.582070827484131
369 -0.577518403530121
370 -0.579150140285492
371 -0.582015752792358
372 -0.582442581653595
373 -0.583370208740234
374 -0.584971189498901
375 -0.588454604148865
376 -0.595993280410767
377 -0.604255437850952
378 -0.609688878059387
379 -0.608756899833679
380 -0.599691987037659
381 -0.589209079742432
382 -0.574401259422302
383 -0.560247421264648
384 -0.546745896339417
385 -0.533863842487335
386 -0.521574735641479
387 -0.508974492549896
388 -0.489413440227509
389 -0.470332443714142
390 -0.452223569154739
391 -0.440128654241562
392 -0.431503921747208
393 -0.423251032829285
394 -0.417380660772324
395 -0.418683737516403
396 -0.418065011501312
397 -0.412524670362473
398 -0.407339960336685
399 -0.402507275342941
400 -0.398024916648865
401 -0.393884003162384
402 -0.389827400445938
403 -0.388083279132843
404 -0.380360752344131
405 -0.372961223125458
406 -0.365878909826279
407 -0.359097570180893
408 -0.345403283834457
409 -0.329335868358612
410 -0.313500285148621
411 -0.297856569290161
412 -0.282376110553741
413 -0.267025887966156
414 -0.25177401304245
415 -0.236592188477516
416 -0.221446797251701
417 -0.206365525722504
418 -0.190379530191422
419 -0.176264435052872
420 -0.166000545024872
421 -0.155637100338936
422 -0.145157590508461
423 -0.134535938501358
424 -0.136438235640526
425 -0.142548397183418
426 -0.148553639650345
427 -0.154478251934052
428 -0.165977463126183
429 -0.183670789003372
430 -0.197871416807175
431 -0.207216516137123
432 -0.211644470691681
433 -0.213905945420265
434 -0.216321989893913
435 -0.21890977025032
436 -0.221685588359833
437 -0.22466678917408
438 -0.227871656417847
439 -0.231319904327393
440 -0.229398742318153
441 -0.217816427350044
442 -0.213048711419106
443 -0.209025323390961
444 -0.205243125557899
445 -0.192398563027382
446 -0.17352207005024
447 -0.154820770025253
448 -0.136259391903877
449 -0.117797553539276
450 -0.0994011610746384
451 -0.0855270624160767
452 -0.0839765071868896
453 -0.0870952308177948
454 -0.0957896634936333
455 -0.10187990218401
456 -0.103136368095875
457 -0.104500561952591
458 -0.105987101793289
459 -0.107610449194908
460 -0.10938511043787
461 -0.114509463310242
462 -0.118677653372288
463 -0.118507653474808
464 -0.11053042113781
465 -0.0991878136992455
466 -0.0869187265634537
467 -0.0698372423648834
468 -0.0441954880952835
469 -0.0171235650777817
470 0.00992438197135925
471 0.0332619398832321
472 0.0489075034856796
473 0.0624779611825943
474 0.0675379037857056
475 0.0727323293685913
476 0.0754447728395462
477 0.0788202732801437
478 0.0823249220848083
479 0.0859581828117371
480 0.0897178202867508
481 0.0947488844394684
482 0.0986246168613434
483 0.0897838622331619
484 0.0808917731046677
485 0.0777370035648346
486 0.0768660753965378
487 0.07356958091259
488 0.068730890750885
489 0.0638711601495743
490 0.0589666068553925
491 0.0539935231208801
492 0.0489274263381958
493 0.0437442660331726
494 0.0384187698364258
495 0.0329263806343079
496 0.0272410213947296
497 0.0213373601436615
498 0.0242457538843155
499 0.0282739698886871
500 0.0382542759180069
501 0.051679715514183
502 0.0672363787889481
503 0.0826101154088974
504 0.0978290885686874
505 0.112922623753548
506 0.127918586134911
507 0.142844513058662
508 0.15772844851017
509 0.172597274184227
510 0.187479764223099
511 0.198233366012573
512 0.207539916038513
513 0.219574227929115
514 0.234370797872543
515 0.251569509506226
516 0.269321918487549
517 0.287224948406219
518 0.303676158189774
519 0.324331223964691
520 0.353611707687378
521 0.37791696190834
522 0.391805768013
523 0.406138837337494
524 0.418304413557053
525 0.429879754781723
526 0.439978808164597
527 0.442148596048355
528 0.444844841957092
529 0.448064148426056
530 0.451804459095001
531 0.456069231033325
532 0.453763633966446
533 0.451524883508682
534 0.450159132480621
535 0.452501833438873
536 0.456516742706299
537 0.464487701654434
538 0.468199819326401
539 0.470115423202515
540 0.468272387981415
541 0.456105589866638
542 0.43970000743866
543 0.428846001625061
544 0.43385323882103
545 0.441721498966217
546 0.449943006038666
547 0.463889122009277
548 0.473317712545395
549 0.489605903625488
550 0.499354422092438
551 0.504105567932129
552 0.50841623544693
553 0.515210628509521
554 0.521027743816376
555 0.521919965744019
556 0.524631142616272
557 0.527815401554108
558 0.526481688022614
559 0.524182558059692
560 0.52233362197876
561 0.520921468734741
562 0.519935667514801
563 0.522623658180237
564 0.5318922996521
565 0.547298550605774
566 0.563202083110809
567 0.582916975021362
568 0.594806253910065
569 0.60354745388031
570 0.612860798835754
571 0.623357534408569
572 0.636380672454834
573 0.650037467479706
574 0.664357125759125
575 0.679371893405914
576 0.692193150520325
577 0.705200791358948
578 0.719939649105072
579 0.726172924041748
580 0.733105480670929
581 0.737008452415466
582 0.746657609939575
583 0.754945516586304
584 0.764074563980103
585 0.774094939231873
586 0.785157203674316
587 0.788314580917358
588 0.789563775062561
589 0.787982702255249
590 0.787331461906433
591 0.78501570224762
592 0.78086131811142
593 0.775472044944763
594 0.768957376480103
595 0.763303637504578
596 0.75848913192749
597 0.754496812820435
598 0.751310586929321
599 0.74891608953476
600 0.746404409408569
601 0.740647256374359
602 0.731861889362335
603 0.718848705291748
604 0.709993004798889
605 0.703964829444885
606 0.698574602603912
607 0.693805038928986
608 0.689639329910278
609 0.686062633991241
610 0.683061242103577
611 0.682113707065582
612 0.681716620922089
613 0.680034458637238
614 0.683194398880005
615 0.69185483455658
616 0.701066195964813
617 0.710853338241577
618 0.721237361431122
619 0.732241809368134
620 0.743897557258606
621 0.753356099128723
622 0.758282840251923
623 0.761184632778168
624 0.76485139131546
625 0.761243581771851
626 0.758304119110107
627 0.756071150302887
628 0.754188895225525
629 0.750739514827728
630 0.747978985309601
631 0.745745301246643
632 0.740196049213409
633 0.738151729106903
634 0.734437942504883
635 0.721391797065735
636 0.708946228027344
637 0.697072565555573
638 0.685734152793884
639 0.674899220466614
640 0.666409730911255
641 0.659018516540527
642 0.652047634124756
643 0.647020101547241
644 0.646292746067047
645 0.641306936740875
646 0.633145451545715
647 0.629978775978088
648 0.628006458282471
649 0.637905418872833
650 0.64971137046814
651 0.661461114883423
652 0.672894835472107
653 0.681360840797424
654 0.690265119075775
655 0.699643194675446
656 0.709517896175385
657 0.719909429550171
658 0.730122327804565
659 0.731026172637939
660 0.729805946350098
661 0.727607429027557
662 0.723508238792419
663 0.719984948635101
664 0.716929912567139
665 0.711102485656738
666 0.711387813091278
667 0.712546825408936
668 0.711394190788269
669 0.703906297683716
670 0.702060461044312
671 0.6937096118927
672 0.685373544692993
673 0.674329280853271
674 0.66111695766449
675 0.648175954818726
676 0.635582506656647
677 0.621722340583801
678 0.607772290706635
679 0.595431923866272
680 0.590494751930237
681 0.586306989192963
682 0.582097291946411
683 0.572680294513702
684 0.564359545707703
685 0.556229650974274
686 0.548262357711792
687 0.540436029434204
688 0.532738626003265
689 0.525151610374451
690 0.517660200595856
691 0.515390753746033
692 0.521979033946991
693 0.526039898395538
694 0.527940511703491
695 0.529484391212463
696 0.531131863594055
697 0.532887578010559
698 0.540550649166107
699 0.550441324710846
700 0.560500621795654
701 0.570734143257141
702 0.587328791618347
703 0.606015503406525
704 0.627649784088135
705 0.647421896457672
706 0.659622371196747
707 0.663759171962738
708 0.664327561855316
709 0.666332125663757
710 0.665703177452087
711 0.662144064903259
712 0.659075140953064
713 0.656510591506958
714 0.65450382232666
715 0.644956350326538
716 0.637812316417694
717 0.628407061100006
718 0.619400560855865
719 0.610829949378967
720 0.599320948123932
721 0.58491849899292
722 0.570837676525116
723 0.565390765666962
724 0.563681662082672
725 0.557121157646179
726 0.554580926895142
727 0.55564296245575
728 0.556963443756104
729 0.558561325073242
730 0.56043553352356
731 0.562600255012512
732 0.565058887004852
733 0.567829728126526
734 0.570912420749664
735 0.574295997619629
736 0.571248710155487
737 0.562982678413391
738 0.5590500831604
739 0.555401861667633
740 0.551200926303864
741 0.547335565090179
742 0.543784976005554
743 0.5405393242836
744 0.540781140327454
745 0.534600973129272
746 0.52724015712738
747 0.520177125930786
748 0.513402223587036
749 0.506907403469086
750 0.507183074951172
751 0.506079614162445
752 0.501393437385559
753 0.503779053688049
754 0.507229745388031
755 0.510929524898529
756 0.514918208122253
757 0.519201934337616
758 0.523788332939148
759 0.528697311878204
760 0.5361168384552
761 0.546993255615234
762 0.558262586593628
763 0.569934844970703
764 0.581652820110321
765 0.587638139724731
766 0.59407639503479
767 0.592819094657898
768 0.588361382484436
769 0.583241641521454
770 0.578675150871277
771 0.574675798416138
772 0.571254253387451
773 0.568394243717194
774 0.566104412078857
775 0.564376950263977
776 0.564441084861755
777 0.566044688224792
778 0.568195700645447
779 0.567429721355438
780 0.566098630428314
781 0.565381646156311
782 0.563449740409851
783 0.561121821403503
784 0.559411883354187
785 0.558313250541687
786 0.557846069335938
787 0.557248711585999
788 0.549509882926941
789 0.536794006824493
790 0.524647951126099
791 0.513130128383636
792 0.502184987068176
793 0.491801559925079
794 0.481956660747528
795 0.469183534383774
796 0.453466981649399
797 0.434865534305573
798 0.416634142398834
799 0.394606173038483
800 0.37694975733757
801 0.357613742351532
802 0.336346030235291
803 0.31933930516243
804 0.30257573723793
805 0.285971283912659
806 0.269455045461655
807 0.253033846616745
808 0.238893389701843
809 0.239210605621338
810 0.240815073251724
811 0.242414981126785
812 0.243979215621948
813 0.245526403188705
814 0.247072994709015
815 0.249542161822319
816 0.256972581148148
817 0.26445746421814
818 0.267643988132477
819 0.271488904953003
820 0.277881383895874
821 0.284377753734589
822 0.291027188301086
823 0.297856986522675
824 0.304905533790588
825 0.312178641557693
826 0.319717407226562
827 0.327548384666443
828 0.335752993822098
829 0.341264516115189
830 0.347176343202591
831 0.35328483581543
832 0.350072950124741
833 0.34405380487442
834 0.338528901338577
835 0.333518981933594
836 0.324771404266357
837 0.315404146909714
838 0.306497097015381
839 0.298023760318756
840 0.289783954620361
841 0.280858188867569
842 0.272318482398987
843 0.264158725738525
844 0.256371229887009
845 0.248939231038094
846 0.241418823599815
847 0.233292549848557
848 0.22480121254921
849 0.216597050428391
850 0.207596197724342
851 0.193780273199081
852 0.179716229438782
853 0.165432617068291
854 0.151629000902176
855 0.140528827905655
856 0.130364894866943
857 0.120309874415398
858 0.110375367105007
859 0.100536182522774
860 0.0907673463225365
861 0.0810540467500687
862 0.0707769095897675
863 0.060267448425293
864 0.0512831509113312
865 0.042601615190506
866 0.033844705671072
867 0.02630265802145
868 0.0207867249846458
869 0.0151189900934696
870 0.00934895128011703
871 0.00348209589719772
872 -0.00255058705806732
873 -0.0092770904302597
874 -0.0161064825952053
875 -0.0229410529136658
876 -0.026027999818325
877 -0.0293865315616131
878 -0.0328797698020935
879 -0.0337124243378639
880 -0.0336991250514984
881 -0.0338804684579372
882 -0.0342306047677994
883 -0.0347404219210148
884 -0.0353644527494907
885 -0.0376547574996948
886 -0.0400391630828381
887 -0.0425557978451252
888 -0.0452096797525883
889 -0.0479496568441391
890 -0.050668865442276
891 -0.0535211935639381
892 -0.056873980909586
893 -0.0609582997858524
894 -0.0651797652244568
895 -0.0692413374781609
896 -0.0739513486623764
897 -0.0810334756970406
898 -0.0871455296874046
899 -0.0936655104160309
900 -0.100354716181755
901 -0.107403978705406
902 -0.11497513204813
903 -0.119276762008667
904 -0.122940726578236
905 -0.124784223735332
906 -0.126100271940231
907 -0.124077767133713
908 -0.12252314388752
909 -0.122654296457767
910 -0.123031847178936
911 -0.122379630804062
912 -0.121918171644211
913 -0.121682487428188
914 -0.120496317744255
915 -0.116924405097961
916 -0.11347234249115
917 -0.11023123562336
918 -0.107013389468193
919 -0.103110007941723
920 -0.0979247540235519
921 -0.0938781648874283
922 -0.0922168344259262
923 -0.0913829058408737
924 -0.0911177098751068
925 -0.0909145176410675
926 -0.0907619372010231
927 -0.0906466245651245
928 -0.0919746086001396
929 -0.0933520346879959
930 -0.0947743877768517
931 -0.0962315648794174
932 -0.0977317988872528
933 -0.0992706418037415
934 -0.100846618413925
935 -0.102305352687836
936 -0.102940306067467
937 -0.105753883719444
938 -0.1083844602108
939 -0.109971679747105
940 -0.110531598329544
941 -0.110971607267857
942 -0.113768227398396
943 -0.116569675505161
944 -0.119502238929272
945 -0.120843060314655
946 -0.122136786580086
947 -0.123502008616924
948 -0.124940037727356
949 -0.126246899366379
950 -0.127479121088982
951 -0.128530085086823
952 -0.129374295473099
953 -0.130303651094437
954 -0.131328508257866
955 -0.132705673575401
956 -0.134300917387009
957 -0.135970622301102
958 -0.137552604079247
959 -0.137714058160782
960 -0.138004913926125
961 -0.1378423422575
962 -0.136926129460335
963 -0.136120572686195
964 -0.135466009378433
965 -0.133261978626251
966 -0.131180971860886
967 -0.129239559173584
968 -0.127433687448502
969 -0.124678388237953
970 -0.121604844927788
971 -0.119491033256054
972 -0.117632433772087
973 -0.115887776017189
974 -0.114217877388
975 -0.11247581243515
976 -0.110605016350746
977 -0.108801275491714
978 -0.107052966952324
979 -0.105356931686401
980 -0.103426650166512
981 -0.10206201672554
982 -0.100691363215446
983 -0.0995453149080276
984 -0.099040612578392
985 -0.0984942838549614
986 -0.097543366253376
987 -0.0971762090921402
988 -0.0970084071159363
989 -0.0968122482299805
990 -0.0969077125191689
991 -0.0969912707805634
992 -0.0970282554626465
993 -0.0970476269721985
994 -0.0970315188169479
995 -0.0969734936952591
996 -0.0968765541911125
997 -0.0967416018247604
998 -0.0965707376599312
999 -0.0963661223649979
1000 -0.0961291566491127
};
\end{axis}

\end{tikzpicture}

%% file: Sections/experiments.tex
\section{Experiments}\label{sec:experiments}
In this section, we begin by executing Algorithm~\ref{alg:BO} to safely tune parameters for a four- and an eight-agent numerical example (Section~\ref{sec:toy}).
Then, we use it to safely tune control parameters for a vehicle platooning simulation (Section~\ref{sec:platooning}).\footnote{
The code is available at \url{https://github.com/tokmaka1/CDC-2025}
}

\subsection{Numerical examples}\label{sec:toy}
For both numerical examples, we optimize over~$T=50$ iterations,  use an RKHS norm upper bound of~$B=1$, and consider the domain~$\domain^N=[0,1]^N$.
For the temporal kernel~\eqref{eq:temporal_kernel}, we use lengthscales~$\ell_\mathrm{RBF}=20$ and~$\ell_\mathrm{Ma12}=5$ and output variances~$\sigma_\mathrm{f,RBF}= \sigma_\mathrm{f,Ma12}=0.1$.
For the spatial kernel, we use~$\ell_\mathrm{Ma52}=0.3$ and~$\sigma_\mathrm{f,Ma52}=1.$

\subsubsection*{Four agents}
\begin{figure}
    \centering
\input{Figures/final/4_agents/reward_development}
    \caption{Reward trajectory of the four-agent toy example.}
    \label{fig:reward-4-agents}
\end{figure}
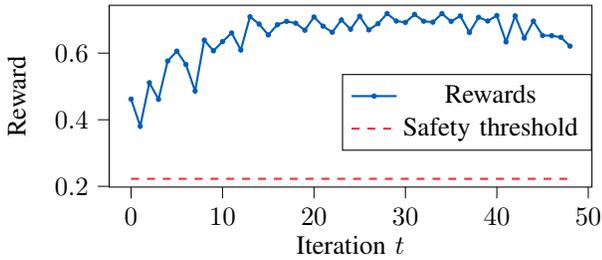

\begin{figure}
\centering
\begin{tikzpicture}[xscale=.735, yscale=.825]
\node[inner sep=0pt] (whitehead) at (4.17, 2.82)
    {\includegraphics[width=6.325cm]{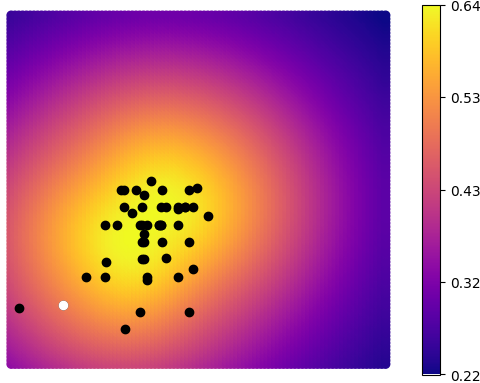}};
\begin{axis}[
tick align=outside,
tick pos=left,
xlabel={$a^{(1)}_{1:t}$},
title={Agent~1},
xmin=0, xmax=1,
xtick style={color=black},
ylabel={$a^{(2)}_{1:t}$},
ymin=0,
ymax=1,
ytick style={color=black}
]
\end{axis}
\end{tikzpicture}
\begin{tikzpicture}[xscale=.735, yscale=.825]
\node[inner sep=0pt] (whitehead) at (4.17, 2.82)
    {\includegraphics[width=6.325cm]{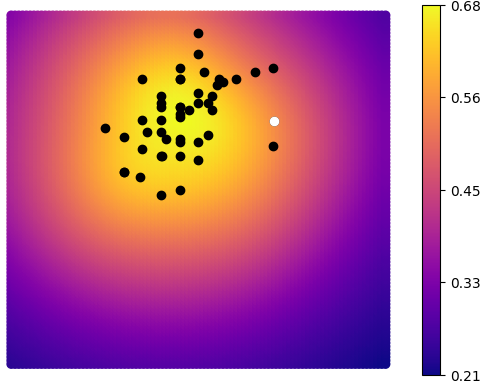}};
\begin{axis}[
tick align=outside,
tick pos=left,
xlabel={$a^{(3)}_{1:t}$},
xmin=0, xmax=1,
title={Agent~4},
xtick style={color=black},
ylabel={$a^{(4)}_{1:t}$},
ymin=0,
ymax=1,
ytick style={color=black}
]
\end{axis}
\end{tikzpicture}
    \caption{\emph{Explored domain of the four-agent toy example.} 
        The black markers represent the sampled control parameters while executing Algorithm~\ref{alg:BO}, whereas the white markers denote the initial safe parameters.
The corresponding upper confidence bounds~$u_t^\A{i}(a_{1:t}^\N{i}, t)$ at iteration~$t=50$ for agents~$i\in\{1,4\}$ are illustrated by the color gradients.
    }
    \label{fig:agents_ucb_4}
\end{figure}

We consider a four-agent distributed MAS with nearest-neighbor communication. 
The agents aim at optimizing an unknown reward function~$f$.
We construct the reward function~$f$ as a member of the pre-RKHS of the Matérn32 kernel with~$\ell_\mathrm{Ma32}=0.4$, RKHS norm~$\|f\|_k=1$, and 1000 center points.
The center points and the coefficients of the pre-RKHS function are sampled uniformly.
The coefficients are then scaled to yield the specified RKHS norm.
Figure~\ref{fig:reward-4-agents} illustrates the reward trajectory over $T=50$ iterations.
The agents clearly improve the reward and stay safe---\ie they do not evaluate parameters that correspond to a reward lower than the pre-defined safety threshold~$h=0.22$, which here is the 20\% quintile of~$f$.
The exploration behavior of the agents is depicted in Figure~\ref{fig:agents_ucb_4}.
Specifically, it shows the sampled parameters~$a_{1:50}^\A{i}$ and the upper confidence bound values~$u_{1:50}^\A{i}$.
The agents explore large parts of the domain and move towards areas with high estimated rewards.

\subsubsection*{Eight agents}
\begin{figure}
    \centering
\input{Figures/final/8_agents/reward_development}
\caption{Reward trajectories for the eight-agent toy example.}
\label{fig:reward-8-agents}
\end{figure}
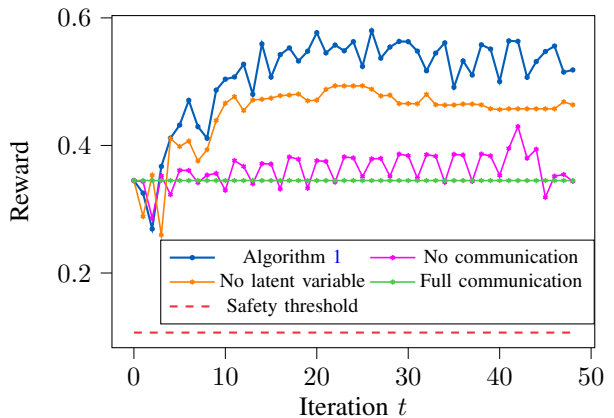

We consider a setup akin to the four-agent example but create the function~$f$ with a lengthscale of~$\ell_\mathrm{Ma32}=0.1$ and use the safety-threshold~$h=0.11$, which corresponds to the 20\% quintile of~$f$.
In this experiment, we compare the performance of our proposed method, \ie Algorithm~\ref{alg:BO}, to other approaches.
Figure~\ref{fig:reward-8-agents} shows the reward trajectory. 
Algorithm~\ref{alg:BO} (blue) shows the most substantial reward improvement while ensuring safety.
In contrast, an approach without time as the latent variable (orange) exhibits an inferior performance as it does not implicitly consider the parameter changes of the non-neighboring agents, which Algorithm~\ref{alg:BO} achieves by leveraging Tools~\ref{tool:time-latent}-\ref{tool:spatio-temporal}.
Moreover, a setting without communication (magenta) also shows a worse performance when compared to Algorithm~\ref{alg:BO}. 
In this setting, the agents implicitly consider the other agents by using time as a latent variable but receive no feedback on the parameter choices of neighboring agents.
Finally, we examine a fully-connected graph, \ie all-to-all communication.
In this setting, each agent models all eight agents.
While this should, in theory, result in the best performance, in practice, we need to choose a coarse discretization of the parameter space due to the high dimensionality.
Therefore, the agents fail to explore. 

\subsection{Vehicle platooning}\label{sec:platooning}
Next, we use Algorithm~\ref{alg:BO} to tune a synchronization controller for vehicle platooning~\cite{besselink2016cyber}.
For the vehicle dynamics, we consider a standard model for heavy-duty vehicles from~\cite{turri2018look}.
We sample the model parameters for the different agents from uniform distributions, thereby creating a heterogeneous MAS.
Specifically, we sample the wheel radius~$r\in [\SI{0.4}{\meter},  \SI{0.6}{\meter}]$, the rolling resistance coefficient~$c_\mathrm{R}\in [\num{4e-3}, \num{8e-3}]$, the cross-sectional area~$A\in [\SI{5}{\meter\squared}, \SI{7}{\meter\squared}]$, the aerodynamic drag coefficient~$C_\mathrm{D}\in[0.4, 0.8]$, and the vehicle's mass~$m\in[\SI{1950}{\kilo\gram}, \SI{2050}{\kilo\gram}]$.
We consider a leader-follower setup, where the first vehicle acts as the leader.
The goal of the other agents is to track a given inter-vehicle distance $d_\mathrm{ref}$.
Each vehicle~$i$ can measure the distance to its preceding vehicle at any time step~$\hat t$, which we denote by~$d_{\hat t}^\A{i}$, and communicate it to its nearest neighbors.
As the leader has no preceding vehicle, it always communicates $d_\mathrm{ref}$.
To achieve synchronization, we leverage a P-controller, as described, for instance, in~\cite[Chapter~4]{lunze2014control}.
That is, each agent~$i$ computes the error
\begin{align}\label{eq:error}
    e^{(i)}_{\hat t} =
    \begin{cases}
        -d^{(i-1)}_{\hat t}  +2d^{(i)}_{\hat t} + d^{(i+1)}_{\hat t}, \quad &\text{if } i> 1, \\
        d^{(i)}_{\hat t} + d^{(i+1)}_{\hat t}, \quad &\text{if } i = 1,
    \end{cases}
\end{align}
and its input as $u^{(i)}_{\hat t} = K_\mathrm{P}^\A{i}e^{(i)}_{\hat t}$.
The error~\eqref{eq:error} has an offset of~$2\cdot d_\mathrm{ref}$ when the inter-vehicle distances equal the reference distance~$d_\mathrm{ref}$.
Hence, with correctly tuned $K_\mathrm{P}^\A{i}$ gains, during perfect synchronization, each vehicle applies a constant input to its motor, which compensates for rolling friction and air resistance, and thus maintains a constant velocity.
We let the leader drive with a constant velocity of $v=\SI{30}{\meter\per\second}$ while the other vehicles tune their individual $K_\mathrm{P}^\A{i}$ parameters to track the reference distance~$d_\mathrm{ref}$. 
We define the reward function as
\begin{equation}\label{eq:platooning_reward}
\begin{aligned}
f(\a_t) = &-\frac{1}{1000 d_\mathrm{ref}N \hat T}\sum_{i=1}^{N} \sum_{\hat t=1}^{\hat T} \lvert d_{\hat t}^\A{i}-d_\mathrm{ref}\rvert \cdot \min_{i,\hat t} d_{\hat t}^\A{i} \\ &-\frac{1}{d_\mathrm{ref}} (d_\mathrm{ref}-\min_{i,\hat {t}}d_{\hat t}^\A{i})\cdot  (1-\min_{i,\hat t}d_{\hat t}^\A{i}),
\end{aligned}
\end{equation}
where~$\a_t$ corresponds to the $K_\mathrm{P}$ gains applied by the four agents at episode~$t$.
The function~\eqref{eq:platooning_reward} rewards a low average deviation from the reference distance~$d_\mathrm{ref}$ and penalizes inter-vehicle distances smaller than~$d_\mathrm{ref}.$
The minimum distance between any vehicle and its preceding vehicle throughout an episode is given by~$\min_{\hat t, i}d_{\hat t}^\A{i}$. 
We use the safety threshold~$h=-1$.
A reward value of~$f(\a_t)=-1$ corresponds to a crash, \ie to~$\min_{\hat t, i}d_{\hat t}^\A{i}=0$ or to a very large deviation from the reference distance~$d_\mathrm{ref}$.

\begin{figure}
    \centering
   \input{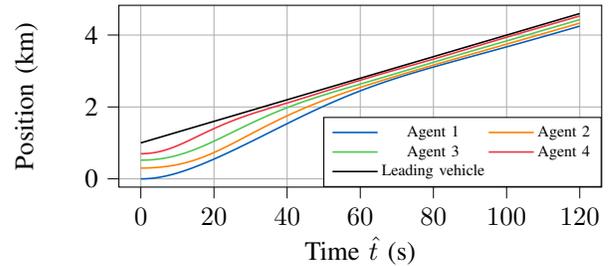}
    \caption{Trajectories of the vehicles using the optimized control gain~$K_\mathrm{P}$.}
    \label{fig:platooning_positions}
\end{figure}

\begin{figure}
    \centering
    \input{Figures/final/reward_development}
    \caption{\emph{Truck platooning.} We tune the gains~$K_\mathrm{P}^\A{i}$ of the proportional controllers of the four following vehicles. We improve the reward and remain safe.}
    \label{fig:platooning}
\end{figure}
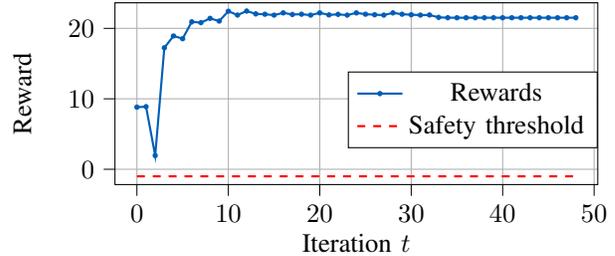

We tune the gains for~$T=50$ episodes with an episode length of~$\hat T=\SI{120}{\second}$ and consider an update interval of~$\Delta t = \SI{0.1}{\second}$.
The vehicles start from the initial positions~$s_0=[\SI{0}{\meter}, \SI{300}{\meter}, \SI{520}{\meter}, \SI{700}{\meter}, \SI{1000}{\meter}]$ with reference distance~$d_\mathrm{ref}=\SI{100}{\meter}$.
For Algorithm~\ref{alg:BO}, we further use~$B=5$, $\domain^N=[0,10]^N$, $\ell_\mathrm{RBF}=20$, $\ell_\mathrm{Ma12}=5$, and $\sigma_\mathrm{f,RBF}= \sigma_\mathrm{f,Ma12}=1$.
For the spatial kernel, we use~$\ell_\mathrm{Ma52}=0.2$ and~$\sigma_\mathrm{f,Ma52}=1.$
We initiate the optimization process with~$\a_0=[4,5,4,5]$ as the initial $K_\mathrm{P}$ gains.
After~$T=50$ iterations, Algorithm~\ref{alg:BO} returns~$K_\mathrm{P}=[6.57, 5.00, 4.44, 6.77]$ as the optimized control parameters that correspond to the largest reward~\eqref{eq:platooning_reward}.
The trajectories of the vehicles using the optimized~$K_\mathrm{P}$ are visualized in Figure~\ref{fig:platooning_positions}.
Moreover, Figure~\ref{fig:platooning} shows the reward over iterations.
Algorithm~\ref{alg:BO} successfully improves the reward without incurring any safety violations, showcasing that our algorithm is applicable to safety-critical real-world scenarios.

%% file: Figures/final/4_agents/reward_development.tex
\begin{tikzpicture}

\definecolor{darkgray176}{RGB}{176,176,176}
\definecolor{steelblue31119180}{RGB}{31,119,180}
\pgfplotsset{
every axis legend/.append style={
at={(1, 0.2)},
anchor=south east,
},
}
\begin{axis}[
tick align=outside,
tick pos=left,
x grid style={darkgray176},
xlabel={Iteration~$t$},
xmin=-2.4, xmax=50.4,
xtick style={color=black},
y grid style={darkgray176},
ylabel={Reward},
ymin=0.197761649638414, ymax=0.743650635331869,
ytick style={color=black},
width=8cm,
height=4cm
]
\addplot [thick, aaltoBlue, mark=asterisk, mark size=1, mark options={solid}]
table {%
0 0.461847126483917
1 0.38074317574501
2 0.511114299297333
3 0.461427360773087
4 0.576298117637634
5 0.606018364429474
6 0.565968811511993
7 0.486363917589188
8 0.639211773872375
9 0.607127130031586
10 0.634645581245422
11 0.660663545131683
12 0.609388887882233
13 0.709347069263458
14 0.687328457832336
15 0.654738962650299
16 0.685506939888
17 0.695378363132477
18 0.690015017986298
19 0.668722987174988
20 0.708702862262726
21 0.680870115756989
22 0.662674903869629
23 0.699713826179504
24 0.67163097858429
25 0.710690319538116
26 0.669785857200623
27 0.688397765159607
28 0.71883749961853
29 0.696390926837921
30 0.692223191261292
31 0.71621960401535
32 0.695829927921295
33 0.692773640155792
34 0.718666851520538
35 0.695336222648621
36 0.711716413497925
37 0.662248909473419
38 0.707470178604126
39 0.696540713310242
40 0.712864637374878
41 0.633817195892334
42 0.711934924125671
43 0.645118892192841
44 0.696276962757111
45 0.653021633625031
46 0.652494847774506
47 0.647573053836823
48 0.620970368385315
};
\addlegendentry{Rewards}
\addplot [thick, aaltoRed, dashed]
table {%
0 0.222574785351753
1 0.222574785351753
2 0.222574785351753
3 0.222574785351753
4 0.222574785351753
5 0.222574785351753
6 0.222574785351753
7 0.222574785351753
8 0.222574785351753
9 0.222574785351753
10 0.222574785351753
11 0.222574785351753
12 0.222574785351753
13 0.222574785351753
14 0.222574785351753
15 0.222574785351753
16 0.222574785351753
17 0.222574785351753
18 0.222574785351753
19 0.222574785351753
20 0.222574785351753
21 0.222574785351753
22 0.222574785351753
23 0.222574785351753
24 0.222574785351753
25 0.222574785351753
26 0.222574785351753
27 0.222574785351753
28 0.222574785351753
29 0.222574785351753
30 0.222574785351753
31 0.222574785351753
32 0.222574785351753
33 0.222574785351753
34 0.222574785351753
35 0.222574785351753
36 0.222574785351753
37 0.222574785351753
38 0.222574785351753
39 0.222574785351753
40 0.222574785351753
41 0.222574785351753
42 0.222574785351753
43 0.222574785351753
44 0.222574785351753
45 0.222574785351753
46 0.222574785351753
47 0.222574785351753
48 0.222574785351753
};
\addlegendentry{Safety threshold}
\end{axis}

\end{tikzpicture}

%% file: Figures/final/8_agents/reward_development.tex
\begin{tikzpicture}

\definecolor{darkgray176}{RGB}{176,176,176}
\pgfplotsset{
every axis legend/.append style={
at={(1, 0.07)},
anchor=south east,
},
}
\begin{axis}[
tick align=outside,
tick pos=left,
x grid style={darkgray176},
xlabel={Iteration~$t$},
xmin=-2.4, xmax=50.4,
xtick style={color=black},
y grid style={darkgray176},
ylabel={Reward},
ymin=0.0829165462404489, ymax=0.603578020259738,
ytick style={color=black},
width=8cm,  
height=6cm,  
legend style={nodes={scale=0.75}, legend columns=2}
]
\addplot [thick, aaltoBlue, mark=asterisk, mark size=1, mark options={solid}]
table {%
0 0.344937562942505
1 0.325192600488663
2 0.269152164459229
3 0.36708727478981
4 0.411423802375793
5 0.43189400434494
6 0.470678091049194
7 0.429431647062302
8 0.411288142204285
9 0.486791670322418
10 0.503858268260956
11 0.507313847541809
12 0.527270913124084
13 0.480119258165359
14 0.559182643890381
15 0.50713449716568
16 0.542316079139709
17 0.552934288978577
18 0.532324850559235
19 0.54770290851593
20 0.576851606369019
21 0.545022547245026
22 0.557376205921173
23 0.548388719558716
24 0.562723696231842
25 0.523883461952209
26 0.579911589622498
27 0.536749839782715
28 0.554283916950226
29 0.56310498714447
30 0.56275475025177
31 0.547887980937958
32 0.517125606536865
33 0.544754087924957
34 0.560904026031494
35 0.491191148757935
36 0.532557666301727
37 0.510331451892853
38 0.557831764221191
39 0.551209032535553
40 0.500168859958649
41 0.56398206949234
42 0.563191294670105
43 0.50680685043335
44 0.531488656997681
45 0.547246396541595
46 0.555819392204285
47 0.514954626560211
48 0.518410980701447
};
\addlegendentry{Algorithm~\ref{alg:BO}}
\addplot [semithick, magenta, mark=asterisk, mark size=1, mark options={solid}]
table {%
0 0.344937562942505
1 0.34341037273407
2 0.285739749670029
3 0.352577328681946
4 0.322839975357056
5 0.361069709062576
6 0.360785394906998
7 0.341629505157471
8 0.35349839925766
9 0.356178194284439
10 0.329752862453461
11 0.376385569572449
12 0.367158621549606
13 0.339969754219055
14 0.37149715423584
15 0.37080979347229
16 0.331723660230637
17 0.381889581680298
18 0.378404021263123
19 0.333028376102448
20 0.376063317060471
21 0.375057727098465
22 0.342634648084641
23 0.382013112306595
24 0.380354434251785
25 0.351089388132095
26 0.378822803497314
27 0.379602164030075
28 0.351717203855515
29 0.386356353759766
30 0.383958756923676
31 0.348444223403931
32 0.385578989982605
33 0.383145779371262
34 0.342276811599731
35 0.386046200990677
36 0.384876161813736
37 0.343725621700287
38 0.386498481035233
39 0.383600562810898
40 0.35315203666687
41 0.395442575216293
42 0.429624229669571
43 0.37997505068779
44 0.394037038087845
45 0.318726778030396
46 0.351774334907532
47 0.35441666841507
48 0.343885838985443
};
\addlegendentry{No communication}
\addplot [semithick, orange, mark=asterisk, mark size=1, mark options={solid} , mark=asterisk, mark size=1, mark options={solid}]
table {%
0 0.344937562942505
1 0.288655310869217
2 0.353422254323959
3 0.259637176990509
4 0.410780429840088
5 0.398295491933823
6 0.40684649348259
7 0.375639945268631
8 0.393400728702545
9 0.439155489206314
10 0.466147392988205
11 0.476472616195679
12 0.454667776823044
13 0.471268087625504
14 0.472384363412857
15 0.474181801080704
16 0.478037655353546
17 0.47893887758255
18 0.480470716953278
19 0.47000578045845
20 0.470986425876617
21 0.487886041402817
22 0.493511348962784
23 0.493264883756638
24 0.493264883756638
25 0.493511348962784
26 0.488174259662628
27 0.477583706378937
28 0.478898346424103
29 0.465619504451752
30 0.465619504451752
31 0.465221673250198
32 0.480118900537491
33 0.463858067989349
34 0.463244378566742
35 0.463390231132507
36 0.464912980794907
37 0.464912980794907
38 0.463603168725967
39 0.457397758960724
40 0.456357032060623
41 0.457397758960724
42 0.457397758960724
43 0.457397758960724
44 0.457397758960724
45 0.457397758960724
46 0.457397758960724
47 0.468391746282578
48 0.463603168725967
};
\addlegendentry{No latent variable}
\addplot [semithick, lightgreen, mark=asterisk, mark size=1, mark options={solid}]
table {%
0 0.344937562942505
1 0.344937562942505
2 0.344937562942505
3 0.344937562942505
4 0.344937562942505
5 0.344937562942505
6 0.344937562942505
7 0.344937562942505
8 0.344937562942505
9 0.344937562942505
10 0.344937562942505
11 0.344937562942505
12 0.344937562942505
13 0.344937562942505
14 0.344937562942505
15 0.344937562942505
16 0.344937562942505
17 0.344937562942505
18 0.344937562942505
19 0.344937562942505
20 0.344937562942505
21 0.344937562942505
22 0.344937562942505
23 0.344937562942505
24 0.344937562942505
25 0.344937562942505
26 0.344937562942505
27 0.344937562942505
28 0.344937562942505
29 0.344937562942505
30 0.344937562942505
31 0.344937562942505
32 0.344937562942505
33 0.344937562942505
34 0.344937562942505
35 0.344937562942505
36 0.344937562942505
37 0.344937562942505
38 0.344937562942505
39 0.344937562942505
40 0.344937562942505
41 0.344937562942505
42 0.344937562942505
43 0.344937562942505
44 0.344937562942505
45 0.344937562942505
46 0.344937562942505
47 0.344937562942505
48 0.344937562942505
};
\addlegendentry{Full communication}

\addplot [thick, aaltoRed, dashed]
table {%
0 0.106582976877689
1 0.106582976877689
2 0.106582976877689
3 0.106582976877689
4 0.106582976877689
5 0.106582976877689
6 0.106582976877689
7 0.106582976877689
8 0.106582976877689
9 0.106582976877689
10 0.106582976877689
11 0.106582976877689
12 0.106582976877689
13 0.106582976877689
14 0.106582976877689
15 0.106582976877689
16 0.106582976877689
17 0.106582976877689
18 0.106582976877689
19 0.106582976877689
20 0.106582976877689
21 0.106582976877689
22 0.106582976877689
23 0.106582976877689
24 0.106582976877689
25 0.106582976877689
26 0.106582976877689
27 0.106582976877689
28 0.106582976877689
29 0.106582976877689
30 0.106582976877689
31 0.106582976877689
32 0.106582976877689
33 0.106582976877689
34 0.106582976877689
35 0.106582976877689
36 0.106582976877689
37 0.106582976877689
38 0.106582976877689
39 0.106582976877689
40 0.106582976877689
41 0.106582976877689
42 0.106582976877689
43 0.106582976877689
44 0.106582976877689
45 0.106582976877689
46 0.106582976877689
47 0.106582976877689
48 0.106582976877689
};
\addlegendentry{Safety threshold}
\end{axis}

\end{tikzpicture}

%% file: Figures/final/reward_development.tex
\begin{tikzpicture}

\definecolor{darkgray176}{RGB}{176,176,176}
\definecolor{steelblue31119180}{RGB}{31,119,180}
\pgfplotsset{
every axis legend/.append style={
at={(1, 0.2)},
anchor=south east,
},
}
\begin{axis}[
tick align=outside,
xmajorgrids,
ymajorgrids,
tick pos=left,
x grid style={darkgray176},
xlabel={Iteration~$t$},
xmin=-2.4, xmax=50.4,
xtick style={color=black},
y grid style={darkgray176},
ylabel={Reward},
ymin=-2.17274007797241, ymax=23.6275416374207,
ytick style={color=black},
height=4cm,
width=8cm
]
\addplot [thick, aaltoBlue, mark=asterisk, mark size=1, mark options={solid}]
table {%
0 8.8159065246582
1 8.88339805603027
2 1.93774616718292
3 17.2436389923096
4 18.9203433990479
5 18.5379180908203
6 20.9278602600098
7 20.8103218078613
8 21.4144306182861
9 21.0356788635254
10 22.4291458129883
11 21.8901958465576
12 22.4548015594482
13 22.0542755126953
14 22.0036277770996
15 21.8768520355225
16 22.199146270752
17 21.9647216796875
18 22.0061779022217
19 21.8770313262939
20 22.1991424560547
21 21.9138031005859
22 21.9852657318115
23 21.8775691986084
24 22.1991176605225
25 22.00559425354
26 21.9335479736328
27 21.8838348388672
28 22.1990127563477
29 22.0055961608887
30 21.9335479736328
31 21.8830528259277
32 21.9030208587646
33 21.5568695068359
34 21.5140056610107
35 21.5097999572754
36 21.5096740722656
37 21.5096702575684
38 21.5096702575684
39 21.5096702575684
40 21.5096702575684
41 21.5096702575684
42 21.5096702575684
43 21.5096702575684
44 21.5096702575684
45 21.5096702575684
46 21.5096702575684
47 21.5096702575684
48 21.5096702575684
};
\addlegendentry{Rewards}
\addplot [thick, red, dashed]
table {%
0 -1
1 -1
2 -1
3 -1
4 -1
5 -1
6 -1
7 -1
8 -1
9 -1
10 -1
11 -1
12 -1
13 -1
14 -1
15 -1
16 -1
17 -1
18 -1
19 -1
20 -1
21 -1
22 -1
23 -1
24 -1
25 -1
26 -1
27 -1
28 -1
29 -1
30 -1
31 -1
32 -1
33 -1
34 -1
35 -1
36 -1
37 -1
38 -1
39 -1
40 -1
41 -1
42 -1
43 -1
44 -1
45 -1
46 -1
47 -1
48 -1
};
\addlegendentry{Safety threshold}
\end{axis}

\end{tikzpicture}

%% file: Sections/conclusions.tex
\section{Conclusions}\label{sec:conclusions}
In this paper, we proposed a BO algorithm for safely tuning control parameters in distributed MAS.
The agents exploit nearest-neighbor communication and explicitly model the influence of the control parameters of neighboring agents on the reward function.
To implicitly account for the behavior of non-neighboring agents, we introduced time as a latent variable.
We also developed a custom spatio-temporal kernel to model the reward as a function of the parameters of neighboring agents and time using GPs.
Our safe BO algorithm leverages these GPs to safely optimize control parameters, which we demonstrated through two numerical examples and on a vehicle platooning simulation.

Potential future work includes proposing a more expressive temporal kernel to include richer prior knowledge when modeling the reward function.
Further, extensions may involve working with different communication (\ie graph) structures, incorporating event-triggered communication protocols, and modifying the framework to multiple safety constraints.